\documentclass[aps,prd,reprint,twocolumn,superscriptaddress,preprintnumbers,nofootinbib]{revtex4-2}

\usepackage{amsthm}
\usepackage{amsmath}
\usepackage{graphicx}
\usepackage{slashed}
\usepackage{amssymb}
\usepackage{soul}
\usepackage{dsfont}
\usepackage{float}
\usepackage[caption=false]{subfig}
\usepackage[colorlinks=True, citecolor=blue, urlcolor=blue, linkcolor=blue]{hyperref}
\usepackage[braket,qm]{qcircuit}
\usepackage{lineno}
\usepackage[dvipsnames]{xcolor}
\DeclareMathOperator*{\argmax}{arg\,max}

\definecolor{darkgreen}{rgb}{0.13,0.55,0.13}
\newcommand*\diff{\mathop{}\!\mathrm{d}}

\newcommand{\nn}{\nonumber}

\newcommand{\be}{\begin{eqnarray}}
\newcommand{\ee}{\end{eqnarray}}

\newcommand{\Tr}{\mathrm{Tr}}

\begin{document}

\title{Liouvillian Dynamics of the Open Schwinger Model: String Breaking and Kinetic Dissipation in a Thermal Medium}

\author{Kyle Lee}
\email{kylel@mit.edu}
\affiliation{Nuclear Science Division, Lawrence Berkeley National Laboratory, Berkeley, California 94720, USA}
\affiliation{Center for Theoretical Physics, Massachusetts Institute of Technology, Cambridge, MA 02139, USA}

\author{James Mulligan}
\email{james.mulligan@berkeley.edu}
\affiliation{Nuclear Science Division, Lawrence Berkeley National Laboratory, Berkeley, California 94720, USA}
\affiliation{Physics Department, University of California, Berkeley, CA 94720, USA}

\author{Felix Ringer}
\email{fmringer@jlab.org}
\affiliation{Thomas Jefferson National Accelerator Facility, Newport News, VA 23606, USA}
\affiliation{Department of Physics, Old Dominion University, Norfolk, VA 23529, USA}
\affiliation{C.N. Yang Institute for Theoretical Physics, Stony Brook University, Stony Brook, NY 11794, USA}
\affiliation{Department of Physics and Astronomy, Stony Brook University, Stony Brook, NY 11794, USA}

\author{Xiaojun Yao}
\email{xjyao@uw.edu}
\affiliation{Center for Theoretical Physics, Massachusetts Institute of Technology, Cambridge, MA 02139, USA}
\affiliation{InQubator for Quantum Simulation, University of Washington, Seattle, WA 98195, USA}

\date{\today}
\preprint{JLAB-THY-23-3894, MIT-CTP-5592, YITP-SB-2023-23, IQuS@UW-21-061}

\begin{abstract}
Understanding the dynamics of bound state formation is one of the fundamental questions in confining quantum field theories such as Quantum Chromodynamics (QCD). One hadronization mechanism that has garnered significant attention is the breaking of a string initially connecting a fermion and an anti-fermion. Deepening our understanding of real-time string-breaking dynamics with simpler, lower dimensional models like the Schwinger model can improve our understanding of the hadronization process in QCD and other confining systems found in condensed matter and statistical systems. In this paper, we consider the string-breaking dynamics within the Schwinger model and investigate its modification inside a thermal medium, treating the Schwinger model as an open quantum system coupled to a thermal environment. Within the regime of weak coupling between the system and environment, the real-time evolution of the system can be described by a Lindblad evolution equation. We analyze the Liouvillian gaps of this Lindblad equation and the time dependence of the system's von Neumann entropy. We observe that the late-time relaxation rate decreases as the environment correlation length increases. Moreover, when the environment correlation length is infinite, the system exhibits two steady states, one in each of the sectors with definite charge-conjugation-parity (CP) quantum numbers. For parameter regimes where an initial string breaks in vacuum, we observe a delay of the string breaking in the medium, due to kinetic dissipation effects. Conversely, in regimes where an initial string remains intact in vacuum time evolution, we observe string breaking (melting) in the thermal medium. We further discuss how the Liouvillian dynamics of the open Schwinger model can be simulated on quantum computers and provide an estimate of the associated Trotter errors.
\end{abstract}

\maketitle

{\tableofcontents}

\section{Introduction} 

Real-time simulations of lattice field theories have recently received significant attention in fundamental nuclear and particle physics. While these simulations pose computational challenges, especially in higher dimensions, recent advancements in quantum computing and error correction~\cite{googleqai1,Sivak_2023} offer the potential to eventually enable large-scale simulations~\cite{Devoret2013, annurev-conmatphys-031119-050605, doi:10.1063/1.5088164, google_supremacy,PhysRevB.107.245423,ibm_faulttol}. The real-time dynamics of field theories can be simulated within the Hamiltonian formulation developed by Kogut and Susskind~\cite{Kogut:1974ag}. Different than in the path integral formulation of lattice field theory that relies on a spatial and temporal lattice discretization, time is kept as a continuous variable within the Hamiltonian formulation and only the spatial directions are discretized. The need to simulate exponentially large Hilbert spaces makes large-scale classical simulations intractable. This necessitates the development of quantum algorithms to simulate for example high energy scattering processes~\cite{Jordan:2011ci,Jordan:2017lea,Bauer:2023qgm,Martin:2023gbo} or field theories at finite chemical potential, which are relevant to nuclear and particle physics~\cite{ Chandrasekharan:1996ih,Martinez:2016yna,Ercolessi:2017jbi,
Dumitrescu:2018njn,Lamm:2018siq,Raychowdhury:2018osk,Roggero:2019myu,Mueller:2019qqj,Avkhadiev:2019niu,Kreshchuk:2020dla,Davoudi:2020yln,Briceno:2020rar,Echevarria:2020wct,Cohen:2021imf,Barata:2021yri,Barata:2022wim,Li:2021kcs,Kharzeev:2020kgc,Bauer:2021gup,Barata:2023clv,Turro:2023xgf}. In particular, lower dimensional lattice field theories that share features with quantum chromodynamics (QCD) have received an increased attention recently. An example is the Schwinger model~\cite{Schwinger:1962tp,Coleman:1975pw}, which corresponds to quantum electrodynamics (QED) in 1+1 dimensions. This U(1) gauge theory coupled to fermions exhibits confinement and chiral symmetry breaking. Besides the similarities with QCD, lower dimensional field theories are an important testing ground for developing simulation protocols in order to eventually build up toward simulations of QCD. Recent work investigated quantum and tensor network simulations of the U(1) gauge theory and the Schwinger model~\cite{PhysRevX.3.041018,Pichler:2015yqa,Muschik:2016tws,Ercolessi:2017jbi,Klco:2018kyo,Kaplan:2018vnj,Magnifico:2019kyj,Butt:2019uul,Chakraborty:2020uhf,Shaw:2020udc,Honda:2021aum,Bauer:2021gek,Nagano:2023uaq,Belyansky:2023rgh,Funcke:2023lli,Angelides:2023bme,PRXQuantum.3.040316,Zhang:2023hzr} and studied Hamiltonian dynamics of non-Abelian lattice field theories~\cite{PhysRevLett.110.125304,PhysRevLett.110.055302,Spitz:2018eps,Klco:2019evd,ARahman:2021ktn,Ciavarella:2021nmj,Ciavarella:2021lel,Yao:2022eqm,Araz:2022tbd,ARahman:2022tkr,Farrell:2022vyh,Farrell:2022wyt,Davoudi:2022xmb,Ciavarella:2022qdx,Zache:2023dko,Hayata:2023bgh,Muller:2023nnk,Cataldi:2023xki,Bauer:2023jvw}, in particular, several efforts aim to understand the thermalization of non-Abelian lattice gauge theory as an isolated quantum system~\cite{Hayata:2020xxm,Hayata:2021kcp,Yao:2023pht}.

\begin{figure}[t]
\includegraphics[scale=0.26]{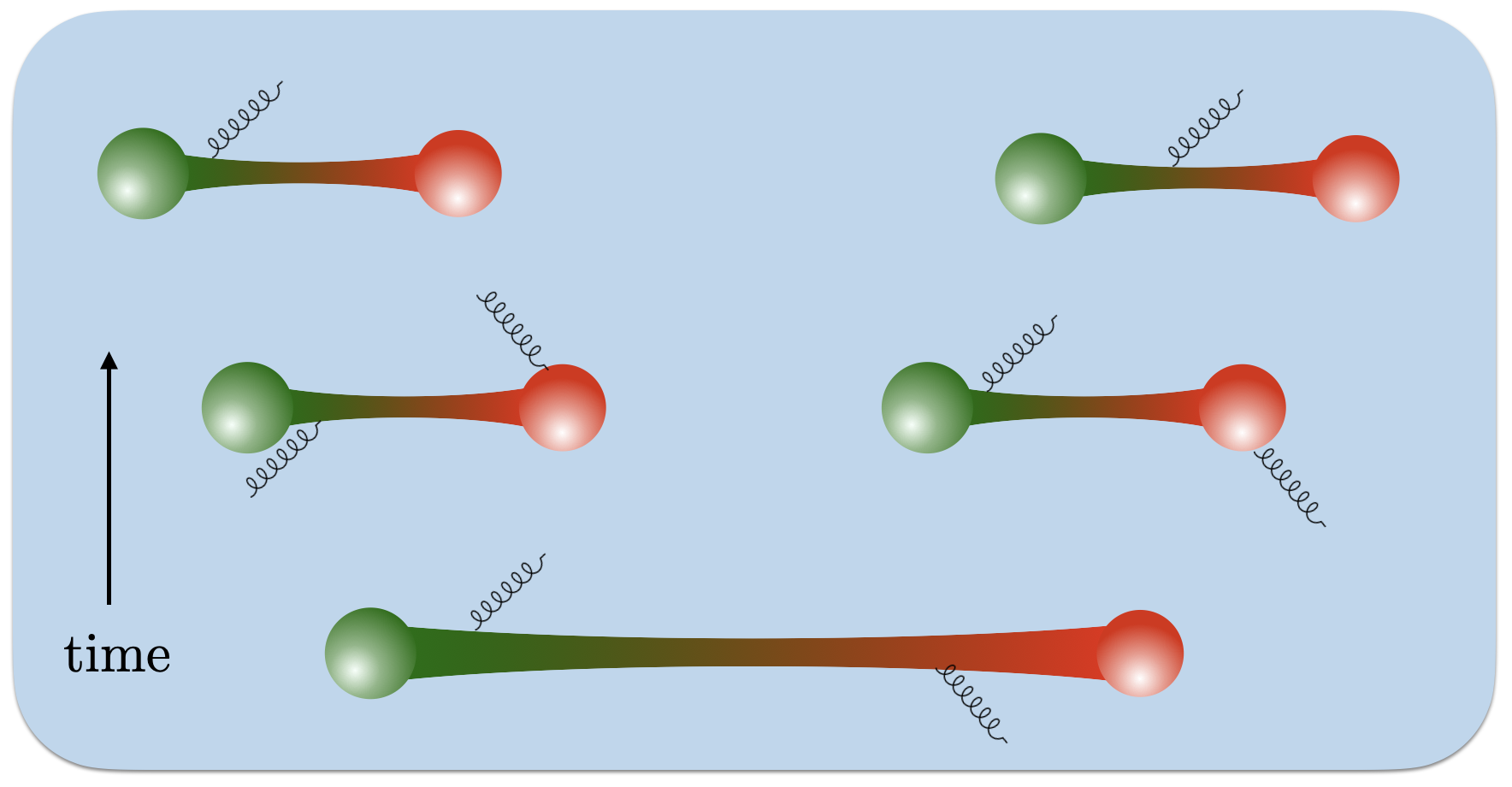}
\caption{Illustration of the string breaking process for the Schwinger model in a thermal medium.~\label{fig:illustration}}
\end{figure}

One of the most intriguing aspects of the Schwinger model is the string-breaking mechanism. This involves a pair of fermion and anti-fermion at a certain distance, connected by a string of electric flux. When the string is sufficiently long, it breaks in real-time, ultimately forming two or more tightly bound fermion anti-fermion pairs, analogous to mesons. The initial energy stored within the string transforms into the kinetic energies of these pairs, which thus separate with corresponding velocities. Details of this mechanism largely depend on the fermion mass and the coupling strength between fermions and the gauge field.

The string-breaking process in the Schwinger model presents fascinating parallels to quark confinement in QCD, where quarks and gluons hadronize into baryons and mesons. This is one of the universe's most compelling enigmas. The phenomenon of string breaking, viewed as a model for hadronization, is also represented in the simulations of high energy particle collisions carried out by Monte Carlo event generators like \textsc{Pythia}\cite{Sjostrand:2007gs}. Additionally, in Ref.\cite{Shen:2022oyg}, initial-state string dynamics and string junctions were found to be necessary for describing particle production in heavy ion collisions within a three-dimensional dynamical initialization model.

Recently, high-energy collider measurements of jet substructure~\cite{Larkoski:2017jix,Marzani:2019hun} have facilitated the direct imaging of the transition between the hadron and parton angular scaling regions, providing a hint of how the confinement scale is set within jets. This can be explicitly observed through measurements of correlations as a function of angle, between the asymptotic energy flux, and is further enhanced by probing these intricate correlations between hadrons with different quantum numbers~\cite{Jaarsma:2023ell,Lee:2023npz,Lee:2023tkr,Lee:2022ige,Komiske:2022enw,Devereaux:2023vjz,ALICE:2023dwg,Tamis:2023guc}. Nevertheless, the theoretical understanding of these measurements, especially in the transition from the universal parton scaling region to the free hadron scaling region—a deeply nonperturbative process—remains a challenging task.

Enhancing our grasp of real-time non-perturbative methods using simpler, lower dimensional models like the Schwinger model could significantly improve our understanding of such real-world collider measurements. Such an understanding could unravel the mystery of quark confinement and has implications for precision measurements of Standard Model parameters~\cite{Hannesdottir:2022rsl}, studies of the quark-gluon plasma (QGP) in heavy ion collisions~\cite{Elfner:2022iae,Busza:2018rrf,Berges:2020fwq}, and the investigation of cold nuclear matter effects at the future Electron-Ion Collider~\cite{AbdulKhalek:2021gbh}. Furthermore, an analogous confinement process occurs in several quasi one-dimensional compounds in condensed matter and statistical systems~\cite{Kormos_2016,Coldea_2010,Lake_2009,Morris_2014,Grenier_2015,Wang_2016}. Hence, studying real-time string-breaking dynamics with the Schwinger model provides a more realistic approach to understanding confinement dynamics in these systems as well.

In this work, we explore the dynamics of the string-breaking mechanism in vacuum and in the presence of a medium, as illustrated in Fig.\ref{fig:illustration}. The static string in the Schwinger model has been studied at both finite temperature and chemical potential~\cite{PhysRevD.19.1188,Pisarski:1982cn,Buyens:2016ecr,Xie:2022jgj} (different lattice field theories at finite temperature and/or chemical potential were also studied in Refs.~\cite{Czajka:2021yll,PhysRevLett.127.057201,Mueller:2021gxd,Schaich:2022duk,Davoudi:2022mbj}). It was observed that the string tension decreases as temperature and/or chemical potential increase. We extend these studies to the dynamical case, where a thermal environment modifies the real-time evolution of the string-breaking process. We find that this environment delays the string-breaking process and reduces the velocity at which the fermion anti-fermion pairs separate. This behavior can be attributed to a quantum drag force acting on the fermion pairs, aligning with findings in the static case. To study real-time dynamics, we consider the Schwinger model interacting with a thermal scalar field via a Yukawa-type coupling. We work in the Brownian motion limit where the environment temperature is high compared to the system's typical energy levels~\cite{Yao:2021lus}. In this limit, memory effects are negligible and the dynamics are Markovian, allowing us to express the evolution of the Schwinger model as an open quantum system in terms of a Lindblad equation~\cite{KOSSAKOWSKI1972247,Lindblad:1975ef,Gorini:1976cm}. The open quantum system framework has been extensively studied for quarkonium dynamics inside the QGP~\cite{Akamatsu:2011se,Akamatsu:2014qsa,Katz:2015qja,Brambilla:2016wgg,Brambilla:2017zei,Blaizot:2017ypk,Kajimoto:2017rel,Blaizot:2018oev,Yao:2018nmy,Akamatsu:2018xim,Miura:2019ssi,Sharma:2019xum,Yao:2020eqy,Akamatsu:2021vsh,Brambilla:2021wkt,Miura:2022arv,Brambilla:2022ynh,Xie:2022tzs,alalawi2023impact}. 

One key aspect of non-equilibrium physics in the open quantum system is the late-time relaxation dynamics toward equilibrium. These relaxation dynamics are governed by the Liouvillian gap, which is given by the eigenvalue of the Liouvillian spectrum whose real part is closest to $0$. This gap is a fundamental quantity of the open quantum system analogous to the energy gap of a Hamiltonian describing a closed quantum system. We determine the Liouvillian spectrum and corresponding eigenmodes of the open Schwinger model for different choices of the environmental correlator (long and short-range correlations), study its dependence on the system size, and compare it to the free fermion model. We find that a long-range correlated environment leads to slower thermalization of the system since the energy and information exchange between the system and environment is slowed when long-range correlations are present in the environment. Moreover, we find that special care needs to be taken in the case of an infinitely long correlation length. In this case, the Liouvillian dynamics of the open quantum system preserve the charge conjugation and parity (CP) symmetry of the system. We decompose the Hilbert space into a CP-even and odd sector. Only in the case of infinite environment correlation, the two sectors evolve independently and there exist two equilibrium states, one in each sector. To study the impact of the environment correlation length on the relaxation dynamics, 
we study the von Neumann entropy of the system that quantifies its decoherence due to the interaction with the environment. These results are closely related to the study and classification of field-theoretical dissipative phase transitions~\cite{doi:10.1142/6020}. Our results provide a starting point for more detailed studies in the future.

Finally, we study the resource requirements for quantum simulations of the Schwinger model as an open quantum system. For this case study, we focus on a quantum algorithm that interleaves short time steps in the system's Hamiltonian evolution with a time evolution operator comprising the Lindblad operators that act on the system and an additional register of ancilla qubits. By using a first-order Trotter decomposition for both unitary operators, we find that, in practice, the Trotter errors associated with the Lindblad evolution may not necessarily increase the total error when compared to the vacuum calculation of the Schwinger model. This is due to some cancellations of errors in the quantum algorithm for simulating the Lindblad evolution, which is an encouraging sign for quantum simulations of open systems in the near to intermediate-term future.

The remainder of this paper is organized as follows. In Section~\ref{sec:schwinger}, we introduce the lattice formulation of the Schwinger model as an open quantum system including the decomposition into separate CP sectors. In Section~\ref{sec:Liouvillian}, we present results for the Liouvillian spectrum and study its relation to the decoherence of the system and relaxation dynamics toward equilibrium. In Section~\ref{sec:StringBreaking}, we present numerical studies of the string breaking process in vacuum and the medium and study its dependence on system parameters. We estimate the Trotter errors of a quantum algorithm for simulating open quantum systems in Section~\ref{sec:QuantumSimulation} and conclusions are drawn in Section~\ref{sec:Conclusions}.

\section{The Schwinger model as an open quantum system \label{sec:schwinger}}

The Lagrangian of the Schwinger model is given by
\begin{equation}
    \mathcal{L}=\bar{\psi}(i \slashed{D}-m) \psi-\frac{1}{4} F^{\mu \nu} F_{\mu \nu}\,,
\end{equation}
with a two-component fermion field $\psi$, the covariant derivative $D_\mu=\partial_\mu-i e A_\mu$, the U(1) gauge field $A_\mu$ and the field strength tensor $F_{\mu \nu}=\partial_\mu A_\nu-\partial_\nu A_\mu$. The Hamiltonian of the Schwinger model can be discretized on a spatial lattice in the axial gauge $A_0=0$ using the staggered fermion formulation and the Jordan-Wigner transform~\cite{Kogut:1974ag}
\begin{align}\label{eq:H_S}
    H_S=
    &\,
    \frac{1}{2a} \sum_{n=0}^{N_{f}-2}\left(\sigma^{+}(n) L_{n}^{-} \sigma^{-}(n+1)+\sigma^{+}(n+1) L_{n}^{+} \sigma^{-}(n)\right)
    \nonumber\\&
    +\frac{1}{2}ae^2 \sum_{n=1}^{N_{f}-1}\ell_{n}^{2}+ \frac{1}{2}m\sum_{n=0}^{N_{f}-1} (-1)^{n} \sigma_{z}(n) \,.
\end{align}
Here $a$ denotes the lattice spacing, and $n$ represents the index of the lattice position $x=na$. Throughout the paper, we will quote parameter values of the Schwinger model in units of $a$.
Moreover, the (anti-)fermion creation/annihilation operators are given by $\sigma^+$/$\sigma^-$ ($\sigma^-$/$\sigma^+$) on even (odd) sites with $\sigma^{\pm}= (\sigma_x \pm i\sigma_y)/2$. Due to the U(1) nature of the theory, we will use electrons interchangeably for fermions and positrons for anti-fermions. Here $L^\pm_n$ correspond to the raising and lowering operators associated with the states of the electric field that lives on the links between lattice sites $n$ and $n+1$. The states of the electric field are labeled by their eigenvalues $e^2\ell_n^2$, which are obtained by acting on these states with the electric field operator squared $e^2E^2(n)$ at site $n$. We assume open boundary conditions, which lead to an unambiguous definition of the environment correlator in the Lindblad equation that will be introduced below.
Under open boundary conditions, the upper limit of the first sum in $H_S$ is $N_f-2$, where $N_f$ is the number of fermion sites. This is twice the number of physical sites $N$ in the stagger fermion formulation so that $N_f=2N$ is an even number. In the case of open boundary conditions, $N_f-1$ gauge links are needed to connect nearest neighbors for $N_f$ fermion sites. 

Physical states have to satisfy Gauss's law, which can be written as
\begin{align}
\label{eqn:gauss}
\ell_{n+1} - \ell_n = - \sigma^+(n) \sigma^-(n) - \frac{(-1)^n - 1}{2} \,.
\end{align}
For the $n=0$ and $n=N_f-1$ sites, imposing Gauss's law requires information about $\ell_0$ and $\ell_{N_f}$, which are not part of the links that we keep track of for dynamics but are determined by the open boundary conditions:
\begin{align}
\ell_0 = 0 \,, \qquad \ell_{N_f} =0 \,.
\end{align}
Other boundary conditions can also be studied, which correspond to cases where the system has a nonzero total charge and/or a uniform background electric field. In one spatial dimension, one can completely integrate out the electric fields by repeatedly using Eq.~\eqref{eqn:gauss}, starting at one end, which leads to non-local interactions between fermions. We will not pursue this here and simply truncate the electric field flux at $|\ell_n|_{\rm max}=1$ for all sites $n$. Studies with higher truncation or electric fields completely integrated out are left for the future. 

\begin{figure}[t]
\includegraphics[scale=0.25]{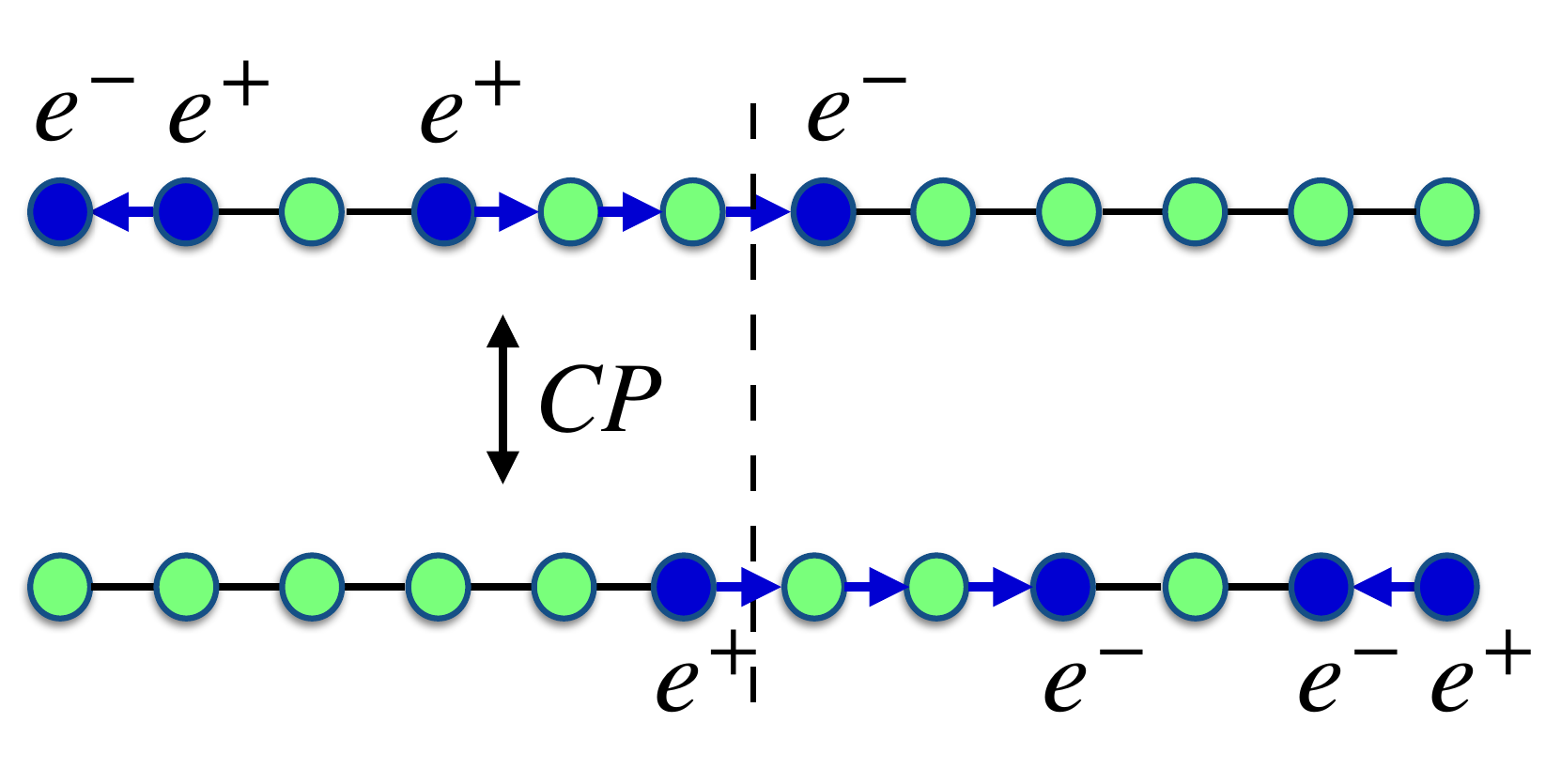}
\caption{Example of how physical states transform under the CP operator. Green (blue) dots are unoccupied (occupied) fermion sites. Fermions (electrons) only live on even sites while anti-fermions (positrons) only live on odd sites. The left and right arrows on the links indicate negative and positive electric fluxes, respectively.~\label{fig:cp}}
\end{figure}

The discretized Hamiltonian has a CP symmetry given by
\begin{align}\label{eqn:cp}
\sigma^\pm(n) &\xrightarrow{CP} \sigma^\mp(N_f-1-n)\,,\nn\\
\sigma_z(n) &\xrightarrow{CP} -\sigma_z(N_f-1-n)\,,\nn\\
L^\pm_n &\xrightarrow{CP} L^\pm_{N_f-2-n}\,,\nn\\
\ell_n &\xrightarrow{CP} \ell_{N_f-2-n} \,.
\end{align}
Under the CP operator, physical states of the theory transform as illustrated in Fig.~\ref{fig:cp}.

Next, we consider the Schwinger model coupled to an environment, which is described by a scalar field theory at thermal equilibrium, as in Ref.~\cite{deJong:2021wsd}. The total Hamiltonian takes the form
\begin{equation}
    H = H_S + H_E + H_I \,,
\end{equation}
where the three terms describe the system, the environment, and their interaction, respectively. The system Hamiltonian $H_S$ is given in Eq.~\eqref{eq:H_S}. The environment Hamiltonian describes a thermal scalar field theory. The interaction Hamiltonian $H_I$ describes the coupling between the Schwinger model and the scalar field theory. Different models of the scalar field interaction terms may be considered.  Here we consider a Yukawa-type interaction 
\begin{align}
H_I = \lambda \int \diff x\, \phi(x) \bar{\psi}(x) \psi(x) \,.
\end{align}
While the system and environment can be strongly coupled, we assume that the interaction between them is sufficiently weak such that the time evolution of the Schwinger model itself is Markovian and a Lindblad equation can be used to describe its time evolution. We consider the quantum Brownian motion limit valid at high temperatures, which allows us to assume that the total density matrix factorizes as $\rho(t)=\rho_S(t) \otimes \rho_E$, where $\rho_S$ denotes the density matrix of the Schwinger model and $\rho_E = e^{-\beta H_E} / \Tr(e^{-\beta H_E})$ is the density matrix of the environment at thermal equilibrium. The Lindblad master equation for $\rho_S$ can be written as~\cite{Yao:2021lus,deJong:2021wsd}
\begin{align}\label{eq:lindblad}
    \frac{\diff \rho_S(t)}{\diff t} 
    & = -i \big[H_S,  \rho_S(t) \big] + a^2 \sum_{x_1,x_2} \, D({x}_1-{x}_2) 
    \nonumber \\ & \times
    \big(L(x_2)\rho_S L^\dagger(x_1) -\frac12 \{L^\dagger(x_1)L(x_2),\rho_S\}\big)\,.
\end{align}
Here $x_1=n_1a$ and $x_2=n_2a$ are discrete spatial coordinates. The environment correlator $D(x)$ only depends on the relative distance between $x_1$ and $x_2$. It can be expressed as
\begin{align}
D(x_1-x_2) = \lambda^2\int_{-\infty}^{+\infty}\!\! \diff(t_1-t_2) \Tr[ \phi(t_1,x_1)\phi(t_2,x_2) \rho_E ] \,,
\end{align}
where $\phi(t,x)$ denotes the scalar field in the interaction picture at thermal equilibrium.  
The Lindblad operators are $L(x) = \bar{\psi}\psi(x) - \frac{1}{4T}[H_S, \bar{\psi}\psi(x)]$ whose notation should be distinguished from the symbol $L_n^\pm$ for the raising and lowering operator of the electric field introduced earlier. On a discrete lattice, we have
\begin{align}\label{eq:LandO}
L(na) &= O(n) - \frac{1}{4T}\left[H_S, O(n)\right] \nn\\
O(n) &= (-1)^n \frac{\sigma_z(n)+1}{2a} \,.
\end{align}
In principle, the environment correlator $D(x)$ can be calculated, which depends on the model for the scalar field theory. For example, for small-size quarkonium inside the QGP, the relevant environment correlator has been formulated~\cite{Brambilla:2017zei,Yao:2020eqy} and studied in both the weak coupling~\cite{Binder:2021otw,Scheihing-Hitschfeld:2022xqx} and strong coupling limits~\cite{Nijs:2023dks,Scheihing-Hitschfeld:2023tuz}. Here instead of calculating the correlator $D(x)$ for a specific scalar field theory model, we directly model the functional form of the correlator. In order to test the dependence of our results on the correlation length of the environment, we use three different models for the correlator: 
\begin{enumerate}
    \item For short-range correlations, we use a delta function: $D_\delta(x)=D_0\delta_{0x}$, where $D_0$ is a constant and $x$ is discrete.
    \item For various intermediate-range correlations, we use a Gaussian\footnote{The Gaussian function decreases much faster at large $x$ than polynomial and exponential functions, which are common functional forms of correlation functions. Studies using polynomial and exponential correlation functions are left for future work.}:
    \begin{equation}\label{eq:D}
       D_G(x) = D_0 \exp\bigg(-\frac{x^2}{2\sigma^2}\bigg) \equiv D_0 G(x,\sigma) \,.
    \end{equation} 
    \item For long-range correlations, we use a constant function $D_c(x)=D_0$.
\end{enumerate}
The normalizations of these three types of correlators are chosen such that they agree at $x=0$. By considering these different choices, we can assess the numerical impact of the environmental correlation length on our results below.

\begin{figure*}[t]
\includegraphics[scale=0.27]{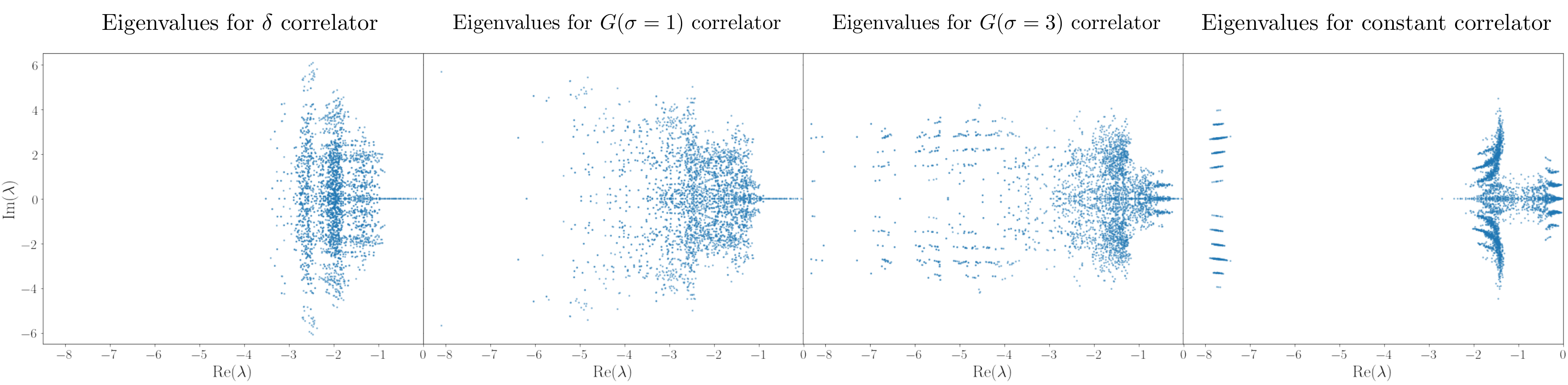}
\caption{Scatter plots of the Liouvillian eigenvalues of the open Schwinger model for $N=4$ lattice sites ($N_f=8$) with $e=0.8,~ m=0.5,~ \beta = 0.1,$ and $ D_0 = 1$ for different types of environmental correlators.~\label{fig:eigenvalues}}
\end{figure*}
Under the CP transformation, the operators $O(n)$ that appear in the Lindblad operators in Eq.~(\ref{eq:LandO}) transform as
\begin{align}
O(n) \xrightarrow{CP} (-1)^n\frac{\sigma_z(N_f-1-n)-1}{2a} \,,
\end{align}
where we have used the fact that $N_f$ is an even number. One can then show that if the environment correlator $D(x)$ is constant, the Lindblad equation given in Eq.~\eqref{eq:lindblad} preserves the CP symmetry. As a result, if an initial state $\rho_S(0)$ is CP-even (odd), the state will remain CP-even (odd) throughout the time evolution. In this case, one can construct two invariant subspaces of the entire Hilbert space: one sector is CP-even and the other one is CP-odd. The construction can be done as follows: We consider each state in the entire Hilbert space. If the state is invariant under the CP transformation, then the state is CP-even. Otherwise, a symmetric linear combination of the original state and the state after the CP transformation leads to a CP-even state while an antisymmetric linear combination yields a CP-odd state.
The CP-even and odd sectors decouple in the time evolution when the environment correlator is constant and thermalize independently. However, we would like to emphasize this is not the case if the environment correlator $D(x)$ is Gaussian or a delta function. An intuitive explanation is as follows: individual Lindblad operator $L(x)$ is not CP invariant. However, when $D(x_1-x_2)$ is constant in the Lindblad equation~\eqref{eq:lindblad}, the two sums over $x_1$ and $x_2$ can be performed independently and then $\sum_x L(x)$ is CP invariant.

\section{Decoherence and relaxation dynamics~\label{sec:Liouvillian}}

The characterization and classification of the relaxation dynamics of open quantum systems  to the steady/thermal state has received significant interest in recent years~\cite{Minganti_2018,Buca:2020vbu,Mori_2020,Longhi_2020,Huybrechts_2020,Hubisz:2020vhx,Haga_2021,de_Leeuw_2021,Zhou_2022,Kawabata_2023,kawabata2022dynamical}. A common approach entails considering either the short or long time non-equilibrium dynamics. At short time scales, the Lindblad evolution can be approximated by a non-hermitian Hamiltonian. In this paper, we primarily focus on long-time relaxation dynamics. The relevance of characterizing these dynamics extends to investigations of non-equilibrium and dissipative phase transitions. For example, while no such transitions occur in the equilibrium state, a phase transition could occur in the decay modes of the Liouvillian. Moreover, the study of dissipative dynamics contributes to understanding phenomena like topological phases, domain walls, non-trivial boundary modes, and exceptional points. While a comprehensive exploration of these aspects within the quantum field theory limit of the Schwinger model is beyond the purview of our current work, we hope this section will provide a valuable starting point for more in-depth future studies.


\subsection{Liouvillian eigenmodes and relaxation dynamics}

We start by rewriting the Lindblad master equation in Eq.~(\ref{eq:lindblad}) in terms of a Liouvillian superoperator $\mathcal{L}$ which operates on the density matrix $\rho$ as
\begin{align}\label{eq:drhoL}
\frac{{\rm d}\rho}{{\rm d}t} \equiv \mathcal{L}\rho\,.
\end{align}
As expected from the open quantum system, the density matrix diagonalizes over time due to thermalization when expressed in terms of the energy eigenstate basis. With our re-expression of the Lindblad equation as a Liouvillian superoperator acting on this density matrix, we are also able to understand \textit{how} the system approaches the thermal state, i.e. the non-equilibrium and relaxation dynamics by carrying out a spectral analysis. That is, we expand the density matrix describing the open quantum system dynamics in terms of eigenmodes of the Liouvillian. The right and left eigenmodes $\rho_j^{R,L}$ are defined as
\begin{equation}
\mathcal{L}\rho_j^R=\lambda_j\rho_j^R\,,\qquad \mathcal{L}^\dagger \rho_j^L=\lambda_j^* \rho_j^L \,,
\end{equation}
where the subscript $j=1,\cdots d^2$ indexes the $j$-th eigenmode with the eigenvalue $\lambda_j$, and $d$ is the size of the Hilbert space. The left and right eigenmodes are orthogonal
\begin{equation}
    \langle \rho_i^L|\rho_j^R\rangle \sim\delta_{ij}\,.
\end{equation}
Here we define the inner product as $\langle A|B\rangle={\rm Tr}[A^\dagger B]$. The dimensionality of the Liouvillian is $d^2\times d^2$, acting on a vectorized density matrix of length $d^2$. In Fig.~\ref{fig:eigenvalues}, we plot the eigenvalues for the open Schwinger model using an $N=4$ lattice with $N_f=8$ fermion sites for different types of interactions that we introduced in Section~\ref{sec:schwinger}. In order to facilitate the visual comparison, we limit the range of $\mathrm{Re}(\lambda_j)$ to $[-8.5,0]$, although the case with a constant environment correlator has a few eigenvalues at much smaller (more negative) real values. These spectra of eigenvalues clearly demonstrate that the non-equilibrium dynamics are nontrivially modified for different types of interactions with the medium. For example, in the case where the interaction corresponds to a delta function for the environment correlator, we observe the emergence of a vertical band structure. This indicates that different subspaces of the Hilbert space decay at separate stages, see also Ref.~\cite{Buca:2020vbu} for example.

Assuming, for now, that there is no degeneracy for the steady state, we can order the eigenvalues such that their real parts are sorted in descending order $0=\mathrm{Re}(\lambda_0)>\mathrm{Re}(\lambda_1)\geq...\geq\mathrm{Re}(\lambda_{d^2-1})$. The time evolution of the general density matrix can then be written, for instance with respect to the right eigenmodes, as
\begin{equation}
\label{eq:evol}
   \rho(t)=\rho_{0} + \sum_{j=1}^{d^2-1}c_j e^{\lambda_j t}\rho_j^R \,.
\end{equation}
The coefficients $c_j$ are obtained by calculating the overlap of the left eigenmodes with the initial state and including an appropriate normalization factor
\begin{equation}
    c_j= \frac{\langle \rho_j^L|\rho(t=0)\rangle}{\langle\rho_j^L|\rho_j^R\rangle}\,.
\end{equation}
This result is obtained by diagonalizing the Liouvillian in Eq.~(\ref{eq:drhoL}). Since the eigenvalues satisfy $\mathrm{Re}(\lambda_{j\geq 1}) < 0$, the density matrix $\rho(t)$ eventually relaxes to $\rho_0$, which is referred to as the (non-equilibrium) steady state, which can be shown to be $1-\frac{H_S}{T}$ for our Lindblad equation~\eqref{eq:lindblad}. It is nothing but the thermal state $e^{-H_S/T}$ in the high-temperature limit, up to corrections of the order $(H_S/T)^2$ (recall that the quantum Brownian motion approximation involves an expansion in $H_S/T$~\cite{Yao:2021lus}). We note that $\rho_0$ is the only eigenmode with a trace equal to $1$, while all the other eigenmodes have vanishing traces. Thus none of the other eigenmodes satisfy the condition to be a density matrix by themselves.

Analyzing the behavior of the open quantum system in terms of the eigenmodes provides means to interpret the non-equilibrium and relaxation dynamics. For example, the approach of the general $\rho(t)$ to $\rho_0$ will be dominated by the first few Liouvillian eigenvalues $\lambda_j$ and the corresponding eigenmodes $\rho_j^R$. In particular, for a given observable $\hat{\cal O}$, the expectation value $\langle\mathcal{\hat{O}}\rangle_{\rho(t)} \equiv {\rm Tr}[\hat{\cal O}\rho(t)]$ will approach the thermal expectation given by the steady-state eigenmode $\langle\mathcal{\hat{O}}\rangle_{\rho_0}$ and its long time rate of approach will be bounded by the real eigenvalue of the first non-stationary eigenmode with smallest $i$ such that $\langle\mathcal{\hat{O}}\rangle_{\rho_i^R} \neq 0$, as $e^{\lambda_i t} \geq e^{\lambda_j t} $ for $i < j$.

In general, the relaxation dynamics towards the stationary state cannot last longer than the rate of decay of the eigenmode $\rho_1^R$. For this reason, it is common to define the Liouvillian or spectral gap $\Delta_1$, which dominates the asymptotic long time decay rate of the Liouvillian, as
\begin{align}
\Delta_1 \equiv |\rm{Re}(\lambda_1)|\,.
\end{align}
The Liouvillian gap $\Delta_1$ is one of the primary features that characterize and are used to classify the dynamics of open quantum systems. In many ways, it is analogous to the spectral gap of closed quantum system Hamiltonians and is associated with the longest lived eigenmode~\cite{Buca:2020vbu}. 

On the other hand, the relaxation time $\tau_R$ is defined as the maximum time at which the following inequality is satisfied~\cite{Haga_2021}
\begin{align}
\text{max}({\tau}):|\langle\mathcal{\hat{O}}\rangle_{\rho(t=\tau)} -\langle\mathcal{\hat{O}}\rangle_{\rho_0} | \geq e^{-1}|\langle\mathcal{\hat{O}}\rangle_{\rho(t=0)} -\langle\mathcal{\hat{O}}\rangle_{\rho_0} |\,,
\end{align}
where the maximization operation is over arbitrary density matrices $\rho(t)$.
Then, from Eq.~\eqref{eq:evol}, one naively expects
\begin{align}
\tau_R \sim \frac{1}{\Delta_1}\,.
\end{align}
This expectation is not always met, and Liouvillian skin effects from boundary conditions are a potential source of deviation from this relation, which were discussed for different quantum mechanical systems in the literature~\cite{Longhi_2020,PhysRevB.102.201103,Haga_2021,Kawabata_2023}. In our case, we do not observe such skin effects, but it would be interesting to study systems with Liouvillian skin effects in the context of quantum field theories as well. 

\begin{figure}[t]
\includegraphics[scale=0.26]{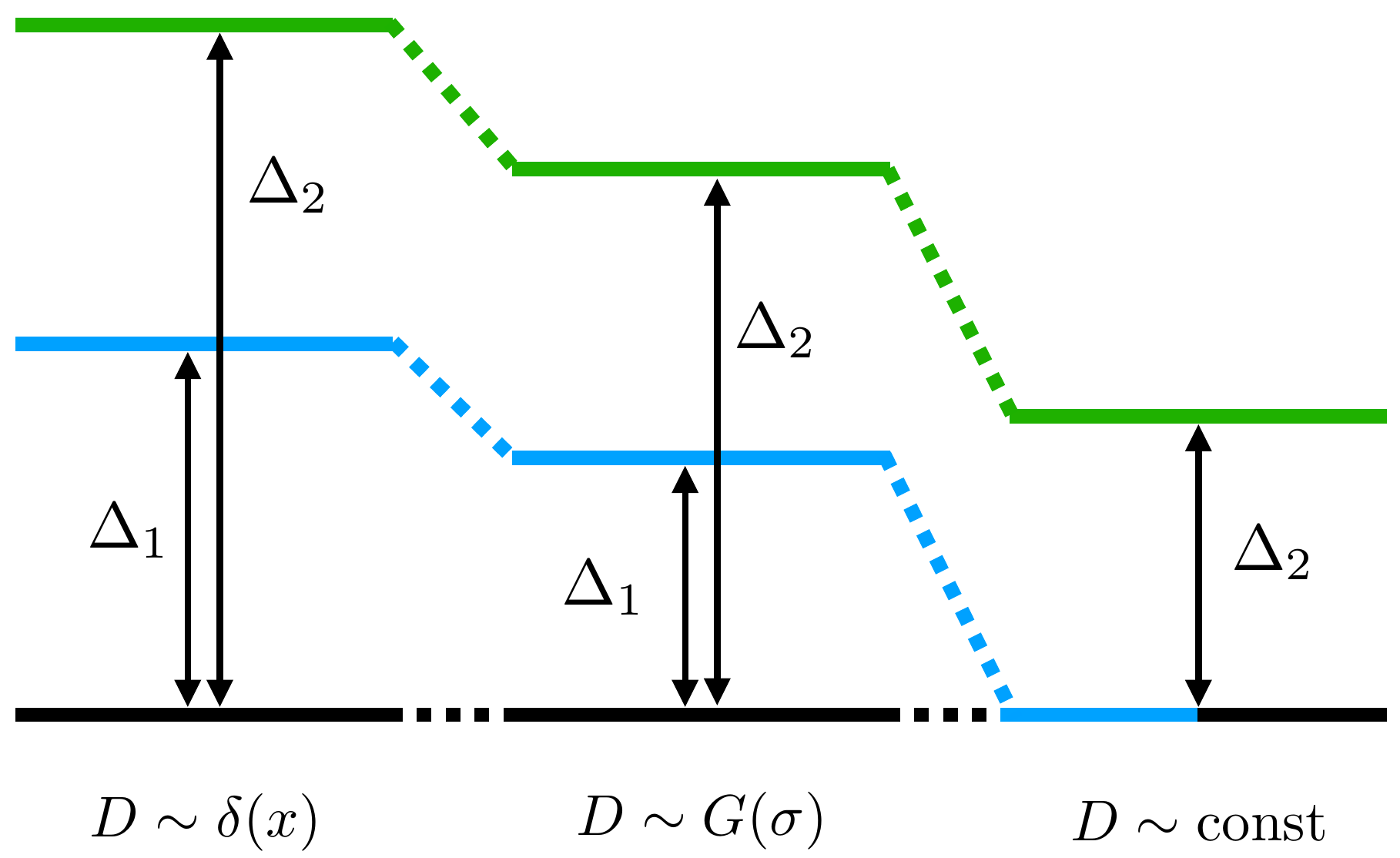}
\caption{Illustration of the gap sizes for different environmental correlators.~\label{fig:DifferentCorrelators}}
\end{figure}

\begin{figure*}[t]
\subfloat[First gap $\Delta_1$.\label{fig:delta1}]{%
  \includegraphics[height=2.4in]{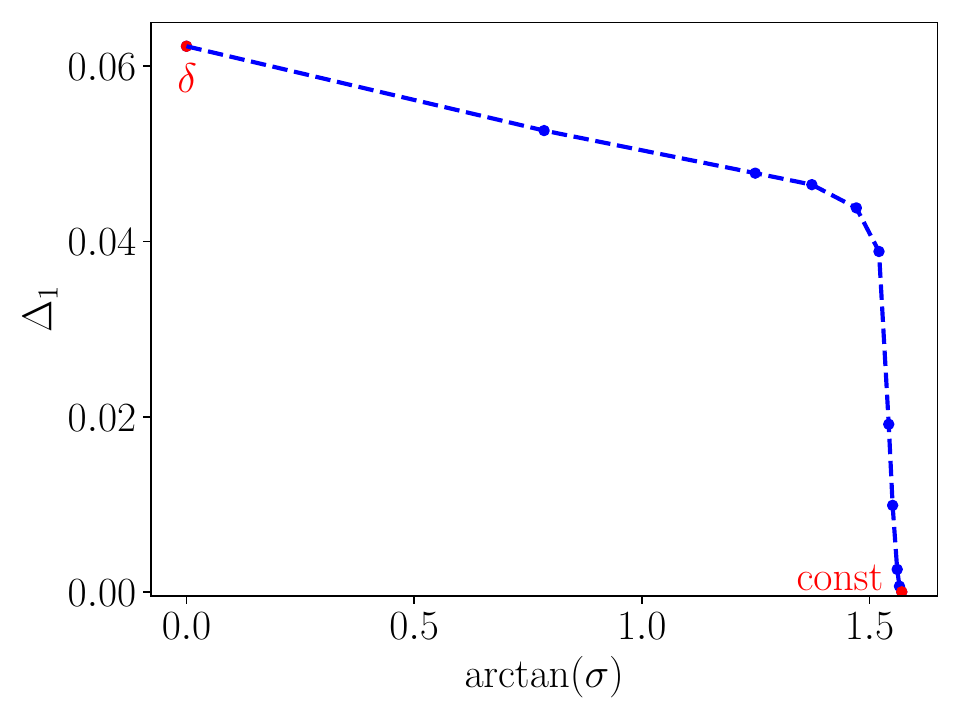}%
}\hfill
\subfloat[Second gap $\Delta_2$.\label{fig:delta2}]{%
  \includegraphics[height=2.4in]{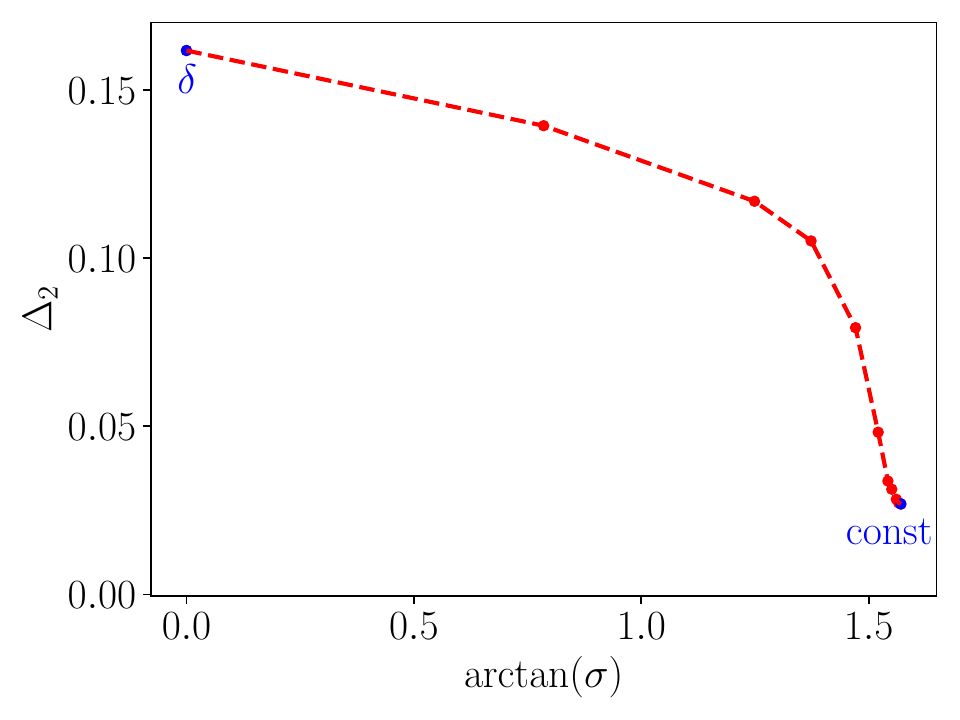}%
}
\caption{The first two Liouvillian gaps, $\Delta_1$ and $\Delta_2$, as functions of the Gaussian width in the environment correlator. We apply the $\arctan$ function to the width to smoothly map both zero and infinite widths onto a finite domain. As indicated in the figure, when the Gaussian width $\sigma$ is zero, the correlator reduces to a delta function, and when $\sigma$ is infinite, the correlator becomes a constant. The $\sigma$ dependence of the relaxation rate is explained in Appendix~\ref{app}.~\label{fig:deltalimit}}
\end{figure*}

As discussed in Section~\ref{sec:schwinger}, the Lindblad equation with a constant environment correlator preserves CP symmetry, leading to the existence of two distinct CP sectors. This implies a degeneracy in the spectrum unless we separate the system into these distinct CP sectors. Therefore, we have two stationary states, $\rho_{0}^{\rm even}$ and $\rho_{0}^{\rm odd}$. That is, our evolution equation in Eq.~\eqref{eq:evol} is now modified to
\begin{align}
   \rho(t)=\,&c_0^e\rho_{0}^{\rm even}+c_0^o\rho_{0}^{\rm odd}\nonumber\\
   &+ \sum_{j=1}^{N_e-1}c_j^e e^{\lambda^e_j t}\rho_j^{R,{\rm even}} + \sum_{k=1}^{N_o-1}c_k^o e^{\lambda^o_k t}\rho_k^{R,{\rm odd}}\,.
\end{align}
Here $N_e$ and $N_o$ are the dimensions of the Hilbert spaces of the CP-even and CP-odd sectors, respectively. They must satisfy the condition $N_e + N_o = d^2$, where $d$ is the dimensionality of the total Hilbert space. While the division between the two CP sectors is clear when the CP symmetry is exact, resulting for example in separate Liouvillian gaps in each sector, it is anticipated that this case will be approximated by a Gaussian environment correlator $D_G(x)$ as its width $\sigma$ increases, even without satisfying the exact CP symmetry. This is illustrated in Fig.~\ref{fig:DifferentCorrelators}, where the constant correlator case $D_c(x)$ depicts the situation before the decomposition into definite CP sectors. As the figure illustrates, the Liouvillian gap denoted by $\Delta_1$ that is present in the case of a delta function and Gaussian correlator reduces as the correlation length of increases. An analytic explanation of this dependence is given in Appendix~\ref{app}. Eventually, the Liouvillian gap vanishes when the correlation length becomes infinite. The vanishing of $\Delta_1$ corresponds to the emergence of two degenerate steady states, one in each CP sector. Consequently, the decay rate towards the global stationary state for $\rho_1^R$ becomes so slow for a Gaussian correlator with a very wide width that its relaxation dynamics at an earlier time scale are primarily dominated by the next gap in the Liouvillian spectrum,
\begin{equation}
\Delta_2\equiv |{\rm Re}(\lambda_2)| \,,    
\end{equation}
which corresponds to the eigenmode $\rho_2^R$. As the width continues to increase, it eventually reaches the limit of a constant environment correlator, where $\rho_1^R$ itself becomes the stationary state in the CP-odd sector. In Fig.~\ref{fig:deltalimit}, we study the behavior of the first Liouvillian gap, $\Delta_1$, and the second gap $\Delta_2$, as functions of the environment correlation length. The Gaussian correlator smoothly connects the cases of a delta function and constant correlator, which have zero and infinite widths, respectively. We found that both gaps demonstrate a smooth behavior while interpolating between the two limits. In particular, we observe that the Lindblad equation with a large-width Gaussian correlator has an approximate CP symmetry, signaled by the vanishing gap $\Delta_1$, which plays an important role in its relaxation dynamics as we will see. We also note that in the infinite correlation length limit, $\Delta_2$ reaches a nonzero value. 
\begin{figure}[t]
\includegraphics[scale=0.5]{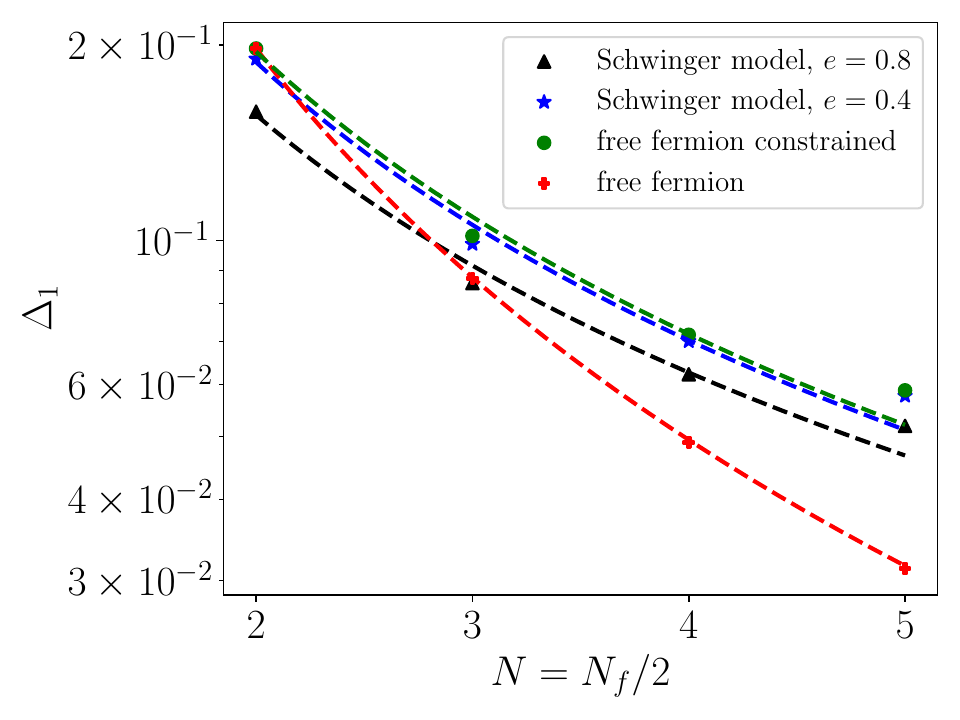}
\caption{The first Liouvillian gap $\Delta_1$ as a function of the number of lattice sites $N$ for the Schwinger model with different couplings $e$ and the free fermion model with the environment correlator described by a delta function in all cases. The $N$ dependence is explained in Appendix~\ref{app}.\label{fig:gap}}
\end{figure}

\begin{figure*}[t]
\includegraphics[scale=0.28]{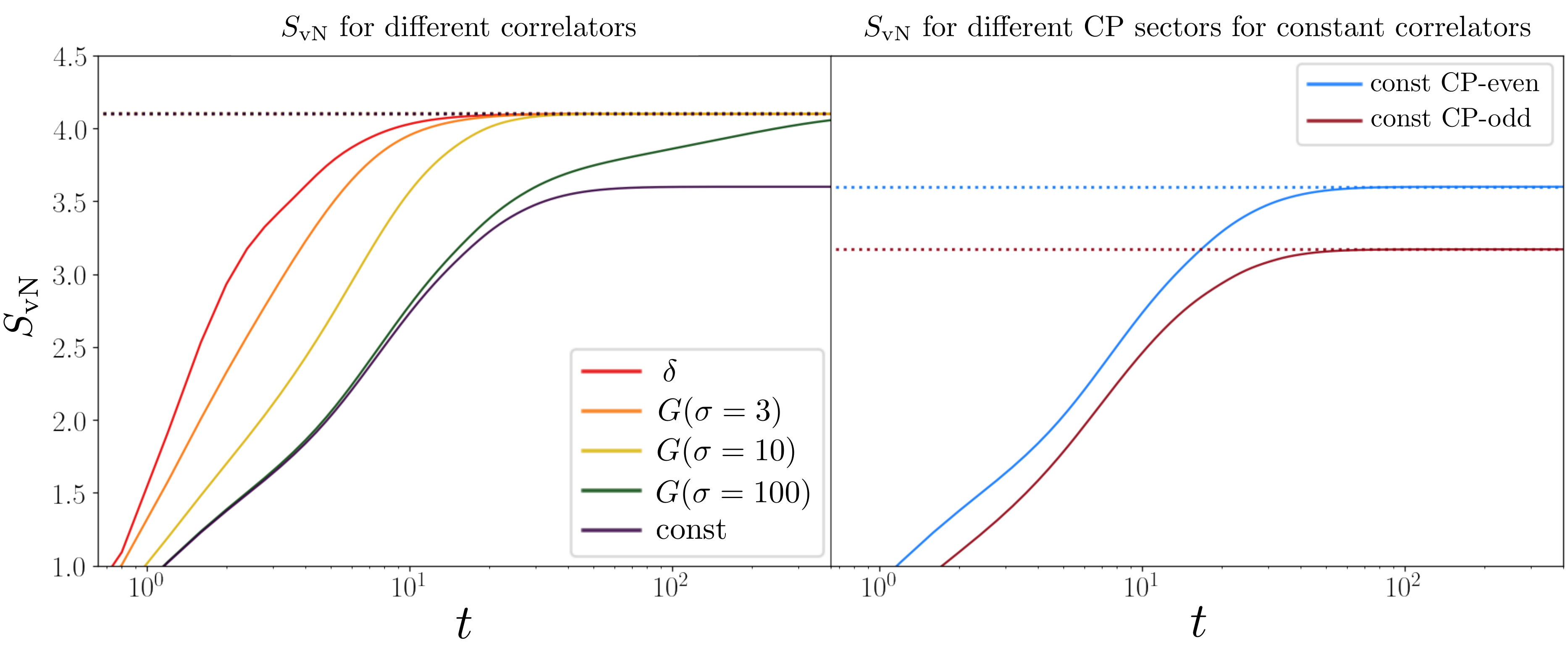}
\caption{The von Neumann entropy $S_{\rm vN}$ of the open Schwinger model for $N=4$ lattice sites with $e=0.8$, $m=0.5$, $\beta=0.1$, and $D_0 = 1$. Left: $S_{\rm vN}$ for different environmental correlators starting from the bare vacuum state in the full Hilbert space as the initial state, which is CP-even. Right: $S_{\rm vN}$ for the constant $D_c=D_0$ environment correlator where the CP sectors are studied separately. In each individual sector we choose appropriate pure states as the initial states. Since the size of the Hilbert space of each sector is smaller than that of the whole system, the maximal von-Neumann entropy of each sector depicted as dashed lines on the right is smaller than the dashed line on the left.~\label{fig:vN}}
\end{figure*}

A further intriguing aspect to explore is the relationship between the Liouvillian gap and system size. With the lattice spacing $a$ fixed, it is expected that the size of the Liouvillian gap decreases as the number of lattice sites increases. This results in a slower thermalization rate since there are more excited modes to equilibrate. However, to obtain the QFT in the continuum limit, one needs to first take $a\to0$ with the system volume fixed, which we leave for future studies. Here we only focus on the case with $a$ fixed. In Fig.~\ref{fig:gap}, we illustrate this phenomenon by plotting the Liouvillian gap $\Delta_1$ for the open Schwinger model with a delta function environment correlator. The figure demonstrates how the gap decreases as the system size increases, which is explained in Appendix~\ref{app}. Also, 
for comparison, we include the case of a 1+1D free fermion theory coupled with a thermal environment with the same delta function correlator. The free fermion theory can be discretized by using the stagger fermion formalism and the Jordan-Wigner transform, which is the non-interacting limit $e\to 0$ of the discretized Schwinger model. The free fermion model Hamiltonian can be mapped onto a spin system analogous to the Schwinger model, which gives
\begin{align}\label{eq:H_S}
    H_{\rm ff}=
    &\,
    \frac{1}{2a} \sum_{n=0}^{N_{f}-2}\left(\sigma^{+}(n) \sigma^{-}(n+1)+\sigma^{+}(n+1) \sigma^{-}(n)\right)
    \nonumber\\&
    + \frac{1}{2}m\sum_{n=0}^{N_{f}-1} (-1)^{n} \sigma_{z}(n) \,.
\end{align}
In order to achieve a direct comparison with the Schwinger model where the total net charge is fixed to zero as a result of the open boundary condition with vanishing electric flux outside the lattice, we also constrain the free fermion system to the sector with zero net charge. The result of $\Delta_1$ for the free fermion case is shown in red in Fig.~\ref{fig:gap}, where the dashed line represents an exact $\propto N^{-2}$ function. The four red points are well described by this function, indicating the first gap $\Delta_1$ in the open free fermion model with a delta environment correlator decreases quadratically with the system size. We note that for other choices of environment correlators, the dependence on $N$ is more complicated than a simple monomial in $N$ but it remains monotonically decreasing with $N$. 

Since we truncate the maximum electric flux at magnitude $1$ for the open Schwinger model, we also need to include a similar constraint for the free fermion model to make a direct comparison. To this end, we only consider states where two neighboring {\it occupied} lattice sites cannot both be electrons or positrons. For example, $|0,e^+,e^-,e^+,e^-,0\rangle$ (where $0$ denotes an unoccupied fermion site) is included in both the constrained and full free fermion models for an $N_f=6$ lattice, whereas $|e^-,0,e^-,e^+,0,e^+\rangle$ is only included in the full free fermion model, as this state would create electric field flux value $2 > |\ell_n|_{\rm max}=1$ at some sites in our constrained Schwinger case. In the second example, the two electrons are on two occupied neighboring sites even though they are separated by one fermion lattice site that is unoccupied. The results for this constrained free fermion model are shown by the green line in Fig.~\ref{fig:gap}. We see that as the coupling in the Schwinger model decreases, the gap results approach those in the constrained free fermion case. The black, blue, and green dashed lines are fits of the form $\propto N^{-\alpha}$. The fitted parameter values are $1.316$, $1.422$ and $1.443$ for the black, blue, and green cases. We see that a monomial in $N$ can approximately describe the $N$ dependence of $\Delta_1$, but not exactly. Given the smooth transition from the constrained Schwinger model to the constrained, free fermion case, we predict that as $e$ decreases, removing the constraint will lead to a convergence towards the free fermion case with the exponent approaching $-2$.

While there are several studies that discuss boundary dissipative systems with bounds on the decay rate of the first Liouvillian gap as a function of the system size~\cite{Prosen_2008,Buca:2020vbu}, a more detailed examination of this phenomenon for the open Schwinger model is left for future work. We now examine the von Neumann entropy of the system, which illustrates that the Liouvillian gap discussed here plays a significant role in describing the relaxation dynamics.

\subsection{Decoherence and von Neumann entropy}

The entropy of quantum systems is frequently studied in the literature. In order to quantify the decoherence of the open Schwinger model, we are going to consider the von Neumann entropy $S_{\rm vN}$, which is given by
\begin{equation}
    S_{\rm vN} = -{\rm tr}[\rho \log \rho] \,.
\end{equation}
The von Neumann entropy vanishes for a pure state where $\rho^2=\rho$ and a finite value for $S_{\rm vN}$ measures the deviation from a pure state. In our case, the decoherence results from the interaction with the thermal environment. The von Neumann entropy is a generalization of the Gibbs (and Shannon) entropy of thermodynamic systems to the quantum case. The phenomenon of decoherence in the density matrix language is frequently discussed in the literature of high energy heavy ion collisions~\cite{Akamatsu:2011se,Blaizot:2018oev,Vaidya:2020cyi} and the concept of entropy has also been discussed in the context of parton distribution functions in Refs.~\cite{Kharzeev:2017qzs,Hagiwara:2017uaz,Zhang:2021hra} and jet physics in Ref.~\cite{Neill:2018uqw}.

In the Schwinger model as an open quantum system, the pure initial state of a string (or analogously the fully unoccupied vacuum state) decoheres due to the interaction with the thermal environment, which is described by the Lindblad equation. Therefore, we obtain a finite value for the von Neumann entropy for $t>0$ which increases as a function of time due to the continued interaction with the environment until the system reaches its steady state. Once the system is in a thermal state, the von Neumann entropy reaches its maximum value, indicating the initial state fully decoheres. The von Neumann entropy is generally bounded by
\begin{equation}
    0\leq S_{\rm vN}\leq \log d\,,
\end{equation}
where $d$ is the dimension of the Hilbert space. As mentioned above, the lower limit is obtained for a pure state, whereas the upper limit is realized for a maximally mixed state proportional to the identity matrix 
$\rho_{\rm mm} = \frac{1}{d} \mathds{1}$. The thermal state generated at late times of the Lindblad evolution approximates the maximally mixed state in the limit $T\to \infty$. Here we explore numerically the real-time dependence of the von Neumann entropy in the Schwinger model as an open quantum system. 

In Fig.~\ref{fig:vN}, we plot the von Neumann entropy starting from an initial pure state as a function of time for $N=4$ lattice sites with parameters $e=0.8,\,m=0.5,\,\beta = 0.1$. In the left panel, we study the time evolution in the full Hilbert space of the Schwinger model (which includes both CP-even and odd sectors) by starting from the bare vacuum state that is CP-even and show the results for different environmental correlators. We observe that the relaxation dynamics significantly depend on the different environment correlation lengths. The von Neumann entropy reaches its maximal value fastest for a delta function correlator $D_\delta$, i.e. for short-range correlated environment that allows for an efficient exchange of momentum and information between the system and environment. For the Gaussian case $D_G\sim G(\sigma)$, we observe that it smoothly approximates the result of a delta function correlator in the limit $\sigma\to 0$. On the other hand, as the correlation length is increased, $S_{\rm vN}$ reaches its maximum value at a much later time. This observation is generally in line with the hierarchy of the gaps for different interactions with $N=4$, as illustrated in Fig.~\ref{fig:deltalimit}. Interestingly, we find that for a constant environmental correlator $D_c$, the von Neumann entropy asymptotes to a lower value. This is due to the decoupling of the two distinct CP sectors in the Liouvillian dynamics. Given that our initial state is CP-even, the thermal state to which it relaxes to is also a CP-even state. The reduction of the Hilbert space for a definite CP sector decreases the final entropy value. We also find that the Gaussian case with a large width does not smoothly approximate the result for a constant correlator $D_c$. While it does approximate it for early times, it eventually deviates and asymptotes to the larger value of $S_{\rm vN}$, as shown by the green line. This is because the Lindblad evolution in this case only approximately preserves the CP symmetry and eventually the entire Hilbert space thermalizes. In this sense, we find a discontinuity of the late time dynamics in the limit $\sigma\to\infty$.

\begin{figure*}[t!]
~~~~
\subfloat[\label{fig:2D-x-t_a}]{%
  \includegraphics[height=2.4in]{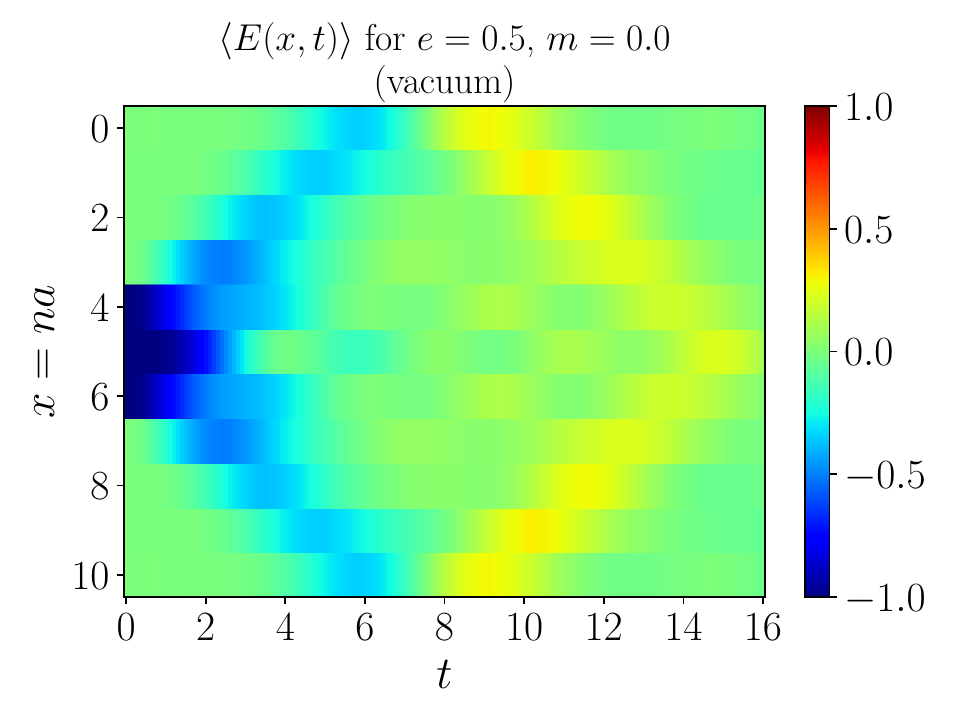}%
}\hfill
\subfloat[\label{fig:2D-x-t_b}]{%
  \includegraphics[height=2.4in]{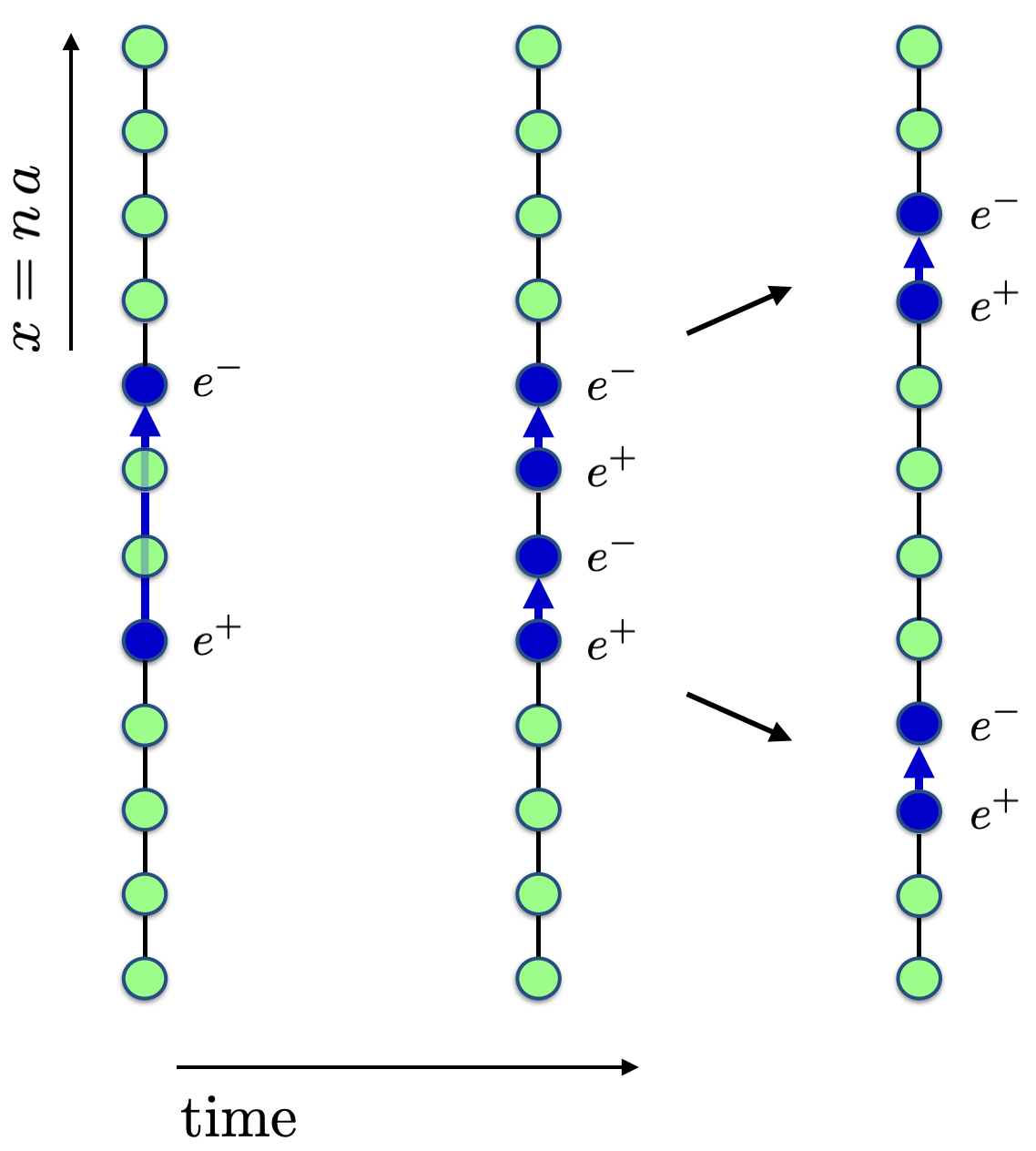}%
}~~~~~~~~~~~~~~~
\caption{String breaking in vacuum for $N_f=12$ fermion lattice sites, corresponding to 11 electric field links. (a): Numerical simulation where the electric field value at each link is used as a measure of the location of the string. (b): Schematic diagram of the string-breaking process. In both cases, the $y$-axis shows the fermion/anti-fermion lattice sites and the $x$-axis shows the time evolution.~\label{fig:2D-x-t}}
\end{figure*}

These results illustrate the nontrivial roles of $\Delta_1$ and $\Delta_2$ as the environment correlation length changes. For instance, the early dynamics of the large Gaussian width case ($\sigma = 100$) closely follow the constant environment correlator case, a similarity that can be attributed to the very close $\Delta_2$ values in both cases. On the other hand, in the case of large Gaussian width, the $\Delta_1$ value is very small, yet nonzero. This contrasts with the constant case, where $\Delta_1$ vanishes, leading to a second stationary state. The small but nonzero $\Delta_1$ value in the large Gaussian width case causes very slow decay, which results in a deviation from the constant case behavior in the large time region. Physically the system quickly thermalizes in the CP-even sector (since the initial state is CP-even), which is governed by $\Delta_2$. The small $\Delta_1$ determines the much slower thermalization between the two CP sectors, which eventually leads to a global thermalization in the whole Hilbert space. This example further demonstrates that non-equilibrium dynamics cannot, in general, be described solely by the first few Liouvillian eigenstates in the full time region. In fact, our example highlights the significant role played by the second gapped state, establishing a principle that can be generalized to higher gapped states in other instances. Non-equilibrium dynamics can display nontrivial behaviors across various time scales, with these behaviors being influenced by multiple eigenstates. For instance, our example clearly exemplifies the type of behaviors that can be expected when a set of eigenvalues are hierarchically separated, i.e., $\cdots \gg \Delta_3 \gg \Delta_2 \gg \Delta_1 $.

In the right panel of Fig.~\ref{fig:vN}, we study the Liouvillian dynamics in each CP sector individually and show the individual von Neumann entropy as a function of time for $D_c$. This clearly demonstrates that initial states from different sectors each relax to their respective thermal states within their sector, which are given by $e^{-H_{e}/T}$ and $e^{-H_{o}/T}$ for the even and odd sectors respectively ($H_e$ and $H_o$ are the corresponding Hamiltonians). In each CP sector, the Hilbert space is smaller than the total Hilbert space with two CP sectors, so the maximum entropy is smaller than the left panel. In addition, we find that the dimensionality of the CP-even Hilbert space is larger compared to the CP-odd case leading to a larger asymptotic value for $S_{\rm vN}$. If one studies the time evolution of an initial state that contains both CP-even and odd parts
\begin{align}
\rho(0) = c \rho^{\rm even}(0) + (1-c) \rho^{\rm odd}(0) \,,
\end{align}
the entropy at late times will be a combination of the two asymptotic entropy values of each CP sector
\begin{align}
S_{\rm vN, asym} = &\ cS_{\rm vN, asym}^{\rm even} + (1-c)S_{\rm vN, asym}^{\rm odd} \nn\\
& -c\log{c} - (1-c)\log(1-c) \,,
\end{align}
where $S_{\rm vN, asym}^{\rm even/odd}$ is the asymptotic entropy value in the CP-even/odd sector.

\section{String dynamics in a thermal medium~\label{sec:StringBreaking}}

In this section, we study the real-time dynamics of the string breaking process in the Schwinger model. 
As mentioned above, the evolution of the string in the Schwinger model can be considered as a model of deconfinement and hadronization in QCD where a quark and an antiquark are separated by a color string, see for example the Lund string model~\cite{Andersson:1983ia}. The in-medium string evolution of the Schwinger model can also be thought of as a model of the quarkonium dynamics in the QGP, where dissociation and recombination of quarkonium occur. String breaking in the Schwinger model has been studied numerically in the vacuum in several previous studies~\cite{Pichler:2015yqa, Magnifico:2019kyj, Honda:2021aum}. We will consider both the vacuum case within our setup as well as for the first time the medium modification to the string breaking dynamics.

\begin{figure*}[t!]
\subfloat[\label{fig:2D-x-t-medium_a}]{%
  \includegraphics[height=1.7in]{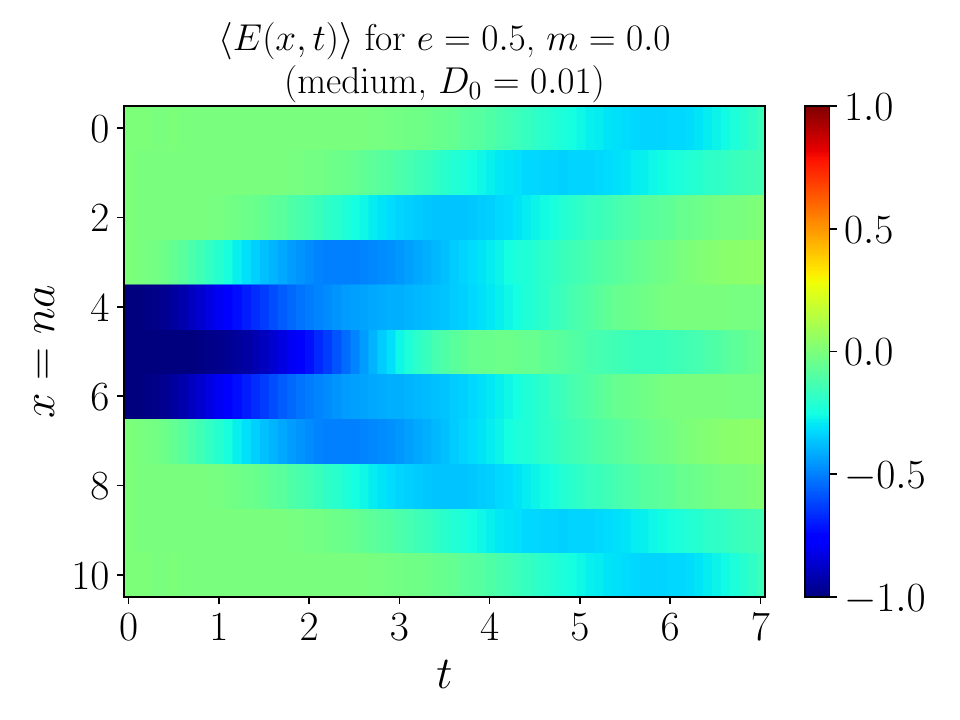}%
}\hfill
\subfloat[\label{fig:2D-x-t-medium_b}]{%
  \includegraphics[height=1.7in]{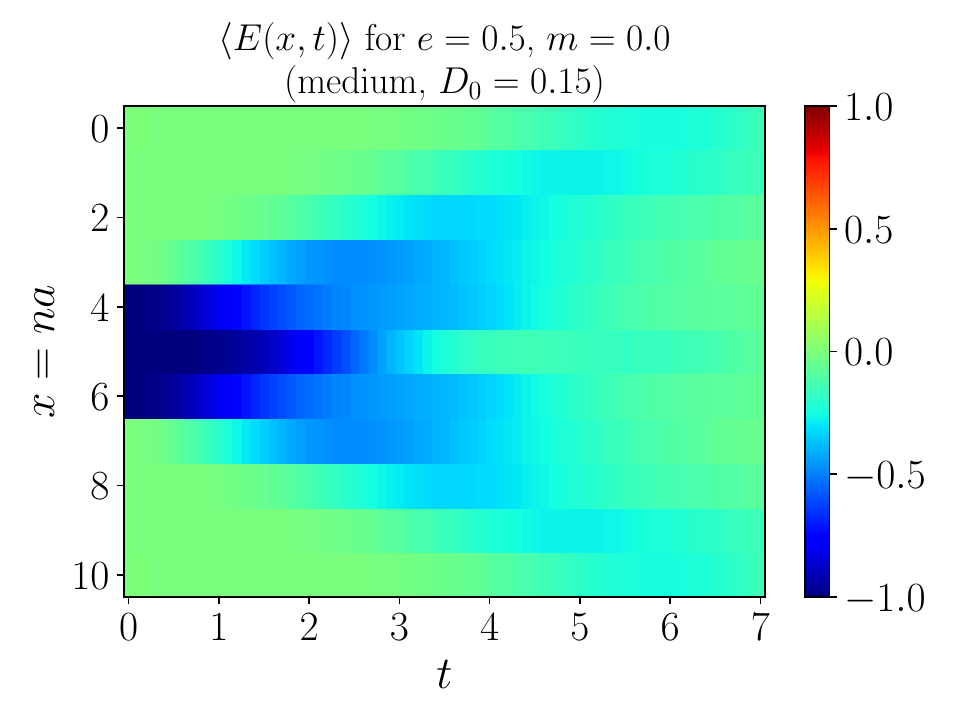}%
}\hfill
\subfloat[\label{fig:2D-x-t-medium_c}]{%
  \includegraphics[height=1.7in]{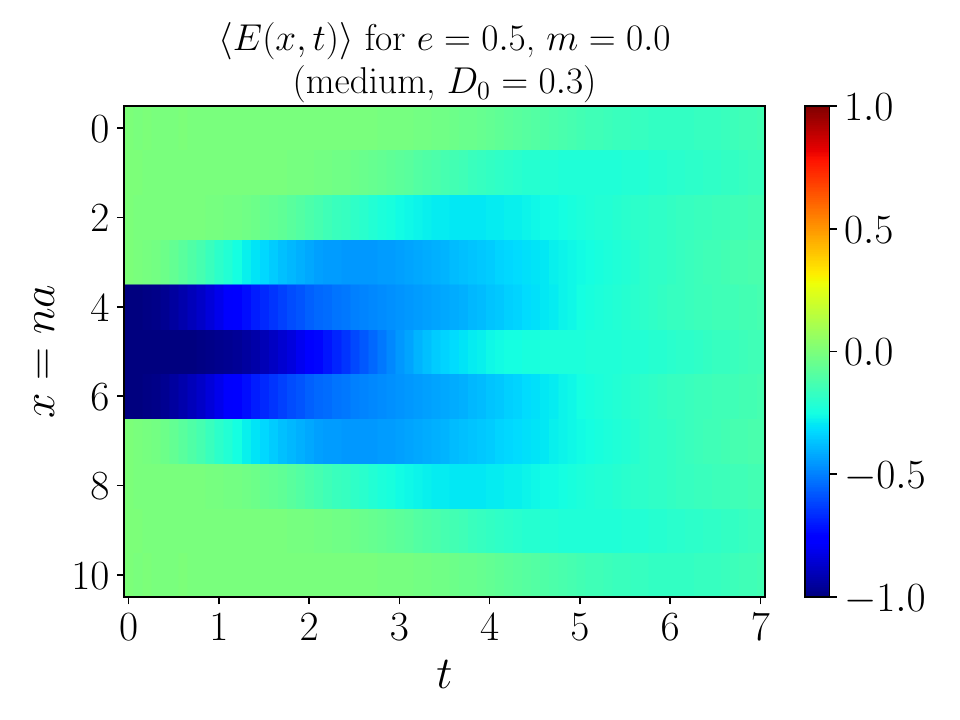}%
}
\caption{String breaking in the medium for $N_f=12$ fermion lattice sites for three different values of $D_0$ with the delta function environment correlator.~\label{fig:2D-x-t-medium}}
\end{figure*}

\subsection{Vacuum}

To begin, we study the string breaking process in vacuum in our setup. We consider an initial configuration where an electron-positron pair separated by some distance is located in the middle of the one-dimensional lattice, connected by a string of electric field links. In order to focus on the dynamics of this string, we will subtract from this configuration the results when a configuration without any fermion and electric flux is initialized, i.e. the bare vacuum state (fully unoccupied). We choose suitable values of $m,e$ where string breaking occurs in the vacuum. In particular, we choose: mass $m=0$, electric charge $e=0.5$, and lattice spacing $a=1$.
Other parameter values will be discussed further in Section~\ref{sect:para}. Since the numerical simulation of the medium case is computationally very expensive we limit ourselves to $N=6$ lattice sites ($N_f=12$ fermion sites) corresponding to $11$ electric field links throughout this section unless stated otherwise. Other numerical approaches such as the quantum trajectory method will allow us to study bigger systems, which will be explored in the future. We note that our initial state corresponds to a bare state where effectively two fermion creation operators are applied to the bare vacuum. It is possible to extend this description and smear the relevant states into wave packets. In this work, we do not pursue this direction further but instead, refer the reader to Refs.~\cite{Surace:2020ycc,Karpov:2020pqe,Milsted:2020jmf}. Another extension is to first prepare the interacting vacuum state (i.e. ground state in energy) and use the state created by applying the fermion creation operators onto the interacting vacuum as the initial state, see e.g. Ref.~\cite{Florio:2023dke}. This is also left for future work.

\begin{figure}[b]
\includegraphics[scale=0.5]{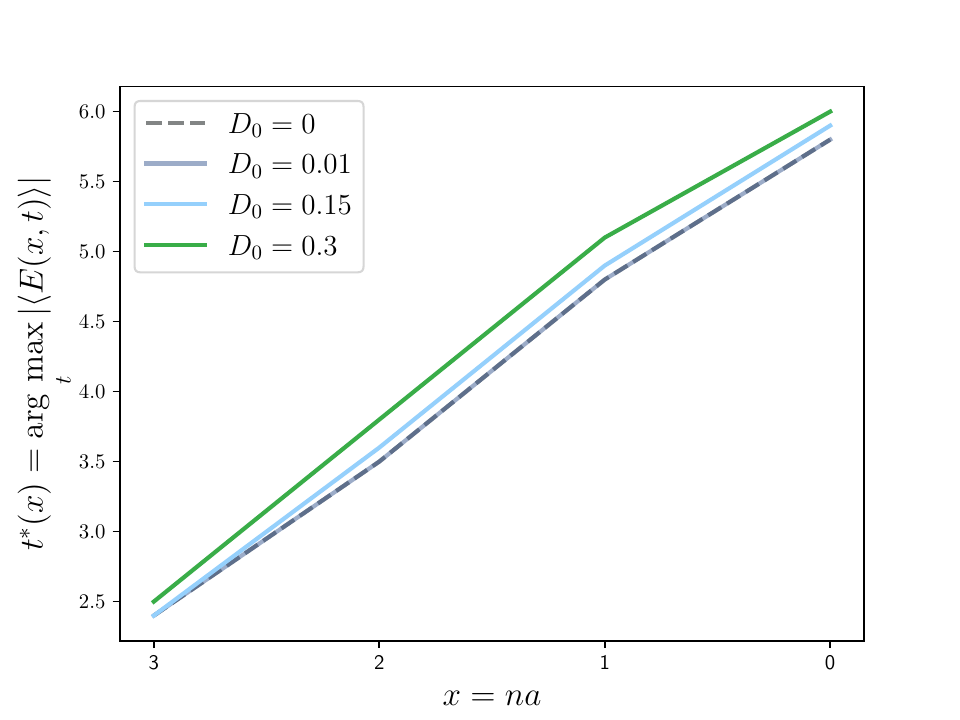}
\caption{The time $t^*(x)$ at which each site $x=na$ reaches the maximum electric field value. We find that $t^*(x)$ is larger for all $x$ in the open system (excluding the trivial central three sites), with a larger delay observed for stronger $D_0$.~\label{fig:t-max}}
\end{figure}

We quantify the presence of the string by measuring the electric field expectation value in units of $e$ as a function of position and real-time, i.e., $\langle E(x,t)\rangle$ (the electric field operator is $eE$). The initial configuration can be seen at $t=0$ on the left end of Fig.~\ref{fig:2D-x-t}.
The string is shown in blue whereas green corresponds to no electric field. 
In our convention, we choose the electric fields pointing upward in the figure to have negative values such that the initial nonvanishing electric fields are $\langle E_n \rangle=-1$. If it is pointing in the opposite direction it will take positive values up to $\langle E_n\rangle=+1$. 

As time evolves, the string breaks\footnote{Again, we will discuss other parameter choices, including a case where the string does not break, in Section~\ref{sect:para}.} and hadronizes into two spatially separated electron-positron pairs (``mesons'') that move away from each other with a certain velocity. These bound meson states can be seen in Fig.~\ref{fig:2D-x-t_a} as small blue regions moving toward the upper and lower edges of the spatial lattice until $t\sim 6$. Eventually, when the two meson states reach the boundary of the lattice, they rescatter and start moving back toward the center of the lattice, as shown by the yellow regions after $t=8$. This is an artifact of the finite size of our setup. With tensor networks, it is possible to simulate significantly larger lattices~\cite{Magnifico:2019kyj}, which we do not pursue in this work.

\subsection{Medium~\label{sec:medium}}

As a starting point, we will first explore how the string breaking process described in Fig.~\ref{fig:2D-x-t} is modified in a thermal medium. The real-time evolution of the string is described by the Lindblad equation given in Eq.~(\ref{eq:lindblad}). For our numerical simulations, we choose the delta function environment correlator $D_\delta=D_0\delta_{0x}$ with different values of the prefactor $D_0$.
Similar to the vacuum case, we again subtract the result of the Lindblad evolved bare vacuum state from the result obtained from an initial string configuration. When $t\gg\frac{1}{|\Delta_1|}$, the initial bare vacuum state also thermalizes and the subtraction gives zero. Therefore, we focus on the time region $t\lesssim\frac{1}{|\Delta_1|}$.

\begin{figure*}[t!]
\subfloat[\label{fig:2D-m-e_vac}]{%
  \includegraphics[height=2.5in]{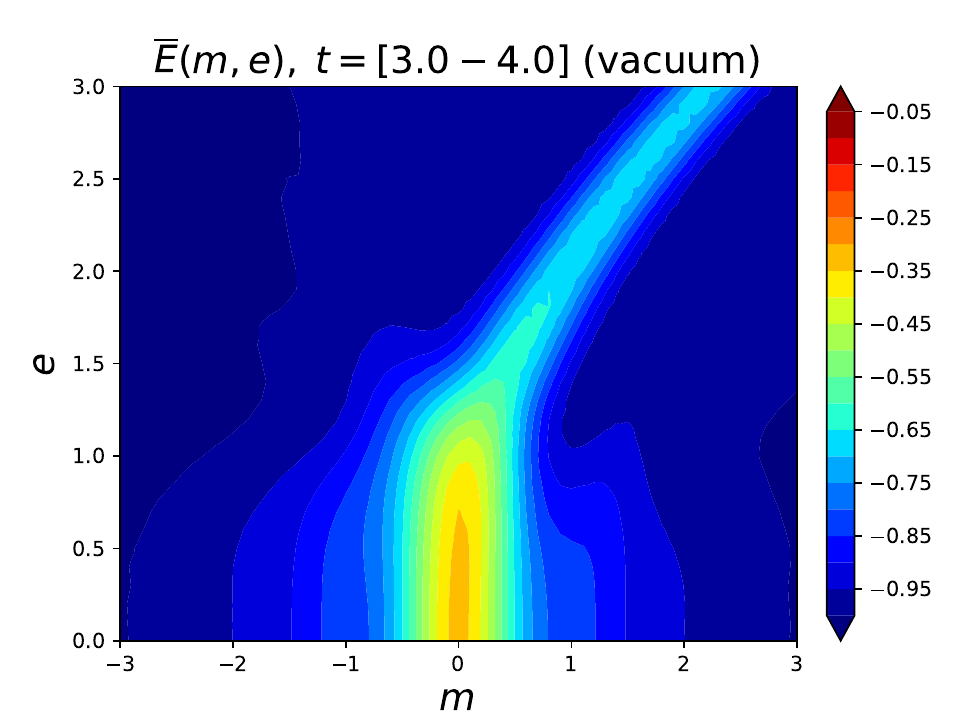}%
}\hfill
\subfloat[\label{fig:2D-m-e_med}]{%
  \includegraphics[height=2.5in]{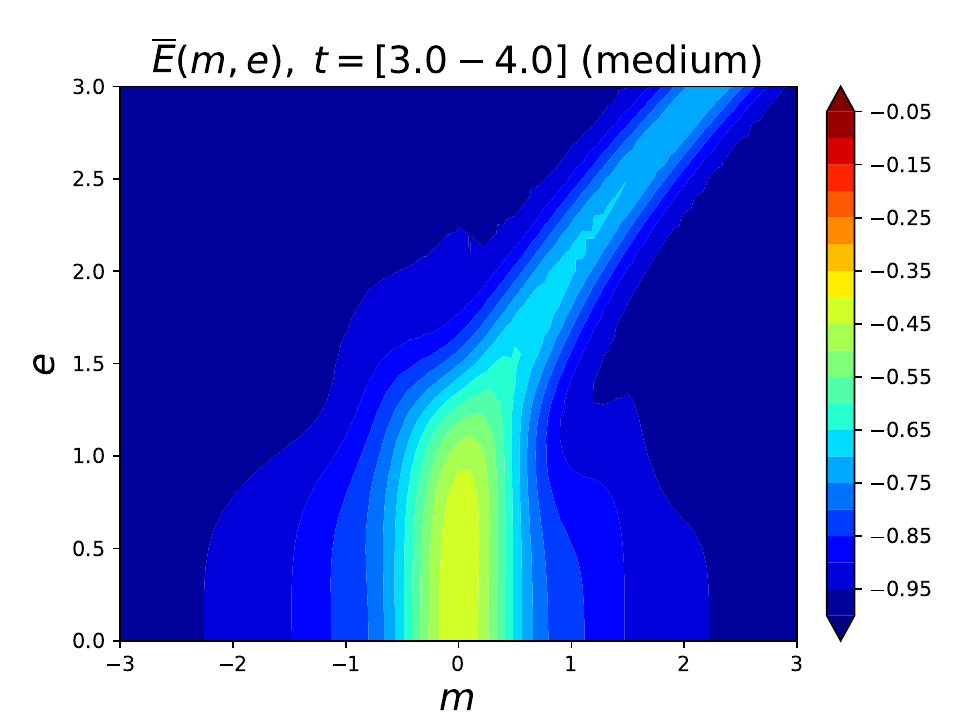}%
}
\caption{String breaking in vacuum (a) and the medium (b) for different values of the mass $m$ and coupling $e$. We show the expectation value of the electric fields in units of $e$ averaged over the three central lattice links and averaged over all the times between $t=3$ and $t=4$ with an initial string located at the three central links. We use a delta function environment correlator with $D_0=0.15$ and $\beta=0.1$ for $N=6$ sites.}
\label{fig:2D-m-e}
\end{figure*}

The open quantum system evolution of the string is shown for an $N=6$ lattice in Fig.~\ref{fig:2D-x-t-medium}. The constant $D_0 = 0.01, 0.15, 0.3$ is increased from left to right. As $D_0$ increases, the system is more significantly modified. Additionally, we investigate whether a delay of the string breaking mechanism is observed as $D_0$ is increased. In order to quantify this effect, we determine the time $t^*(x)$ at which each site $x=na$ reaches its maximum electric field value,
\be
t^*(x) = \argmax_{t} |\langle E(x,t)\rangle| \,.
\ee
Here $t$ is chosen in an interval $t \in [0,t_{\rm max}]$, where $t_{\rm max}$ is determined by the onset of boundary effects due to the finite size of the lattice and is roughly $t\simeq 6$ as shown in Fig.~\ref{fig:2D-x-t}. The results are shown in Fig.~\ref{fig:t-max}, where we plot $t^*$ as a function of the site $x=na$ with the index $n\in[0,1,2,3]$ (we excluded the middle sites where the string is initialized). We see that in the limit of small $D_0$ ($D_0=0.01$) the open system behavior approaches the vacuum behavior. We also find that $t^*(x)$ is larger for all $x$ in the open system, with a longer delay observed for a larger $D_0$. This delay can be understood from the medium dissipation effect, which is already known in the quarkonium dynamics in a thermal medium~\cite{Akamatsu:2018xim,Miura:2019ssi}. When the initial pair of the electron and positron is broken into two mesons, the energy stored in the initial electric string is converted into the masses of the extra two fermions and the kinetic energies of the two mesons. The initial kinetic energies are the same as in the vacuum evolution. The dissipative term, i.e., the $H_S/T$ term in the Lindblad operators shown in Eq.~\eqref{eq:LandO}, reduces the kinetic energy of the system and plays a crucial role for the system to approximately thermalize. As a result of the kinetic dissipation, the velocities at which the two mesons move away from each other decrease, and the separation of the two mesons is delayed.

We also note that for a sufficiently long time, the string magnitude at every site tends to zero. This is because we take the difference between the result obtained from an initial bare string state and that from an initial bare vacuum state. After a long time, the system reaches the approximate equilibrium state, which is the same for different initial conditions in the case of the delta function environment correlator. We would like to point out that if the initial state was prepared by applying two fermion creation operators on the interacting vacuum state, the subtraction performed here would not be necessary.

\subsection{Dependence on system parameters 
~\label{sect:para}}
The real-time dynamics of the string breaking depend on the fermion mass $m$ and coupling $e$.
In the vacuum, there are three different regimes~\cite{Pichler:2015yqa,Magnifico:2019kyj}, which are quantified in Fig.~\ref{fig:2D-m-e_vac} with a metric $\overline{E}$ defined as the average expectation value of the electric fields in units of $e$ at the three central sites over a specified time window:
\be
\overline{E} \equiv \frac{1}{3(t_2-t_1)} \int_{t_1}^{t_2} {\rm d}t \sum_{n\in[4,5,6]} \langle E(na, t) \rangle 
\ee
The three regimes are as follows:
\begin{figure*}[t!]
\subfloat[\label{fig:string_a}]{%
  \includegraphics[height=1.7in]{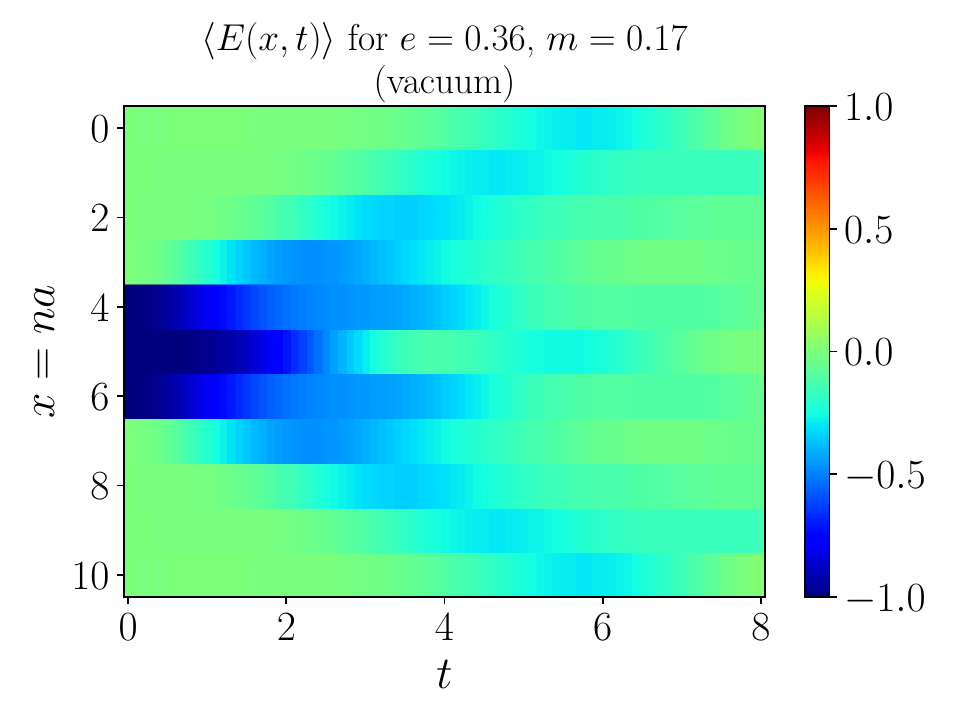}%
}\hfill
\subfloat[\label{fig:string_b}]{%
  \includegraphics[height=1.7in]{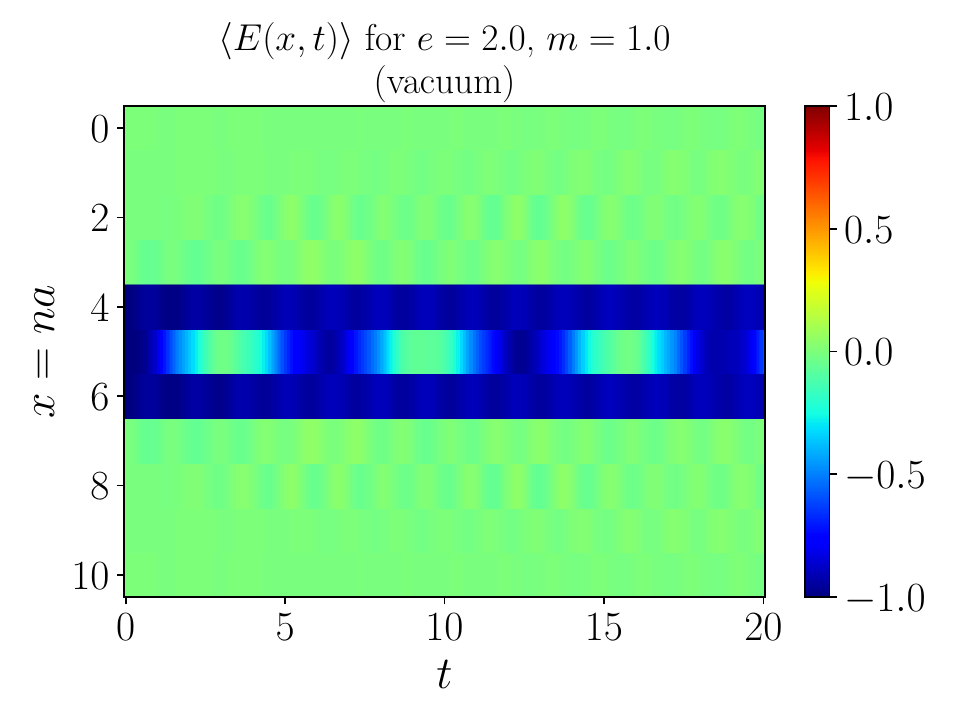}%
}\hfill
\subfloat[\label{fig:string_c}]{%
  \includegraphics[height=1.7in]{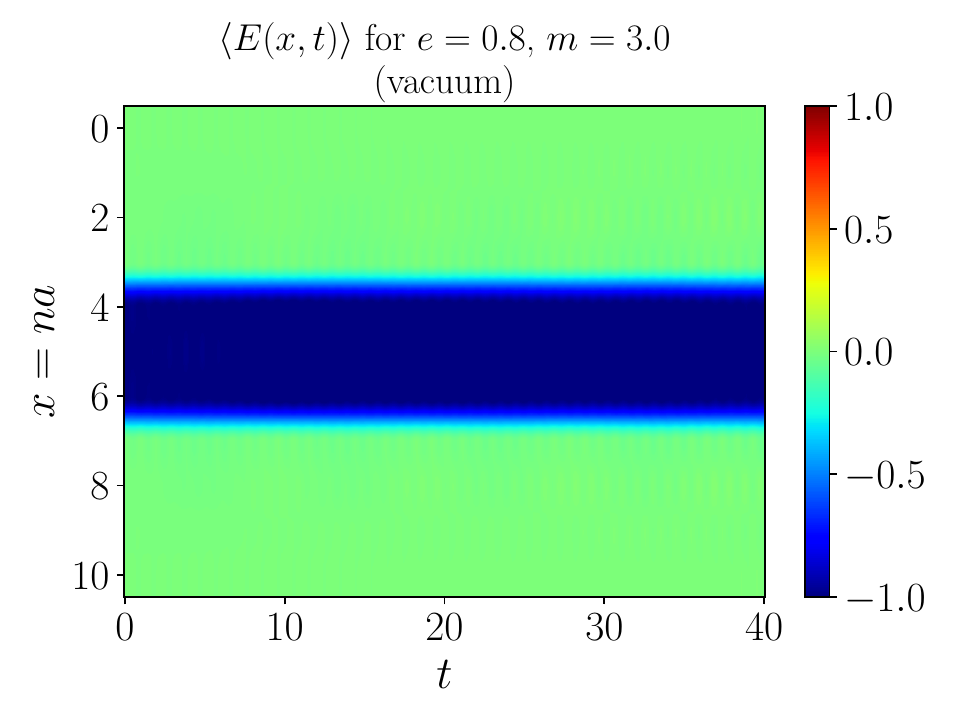}%
}

\subfloat[\label{fig:string_d}]{%
  \includegraphics[height=1.7in]{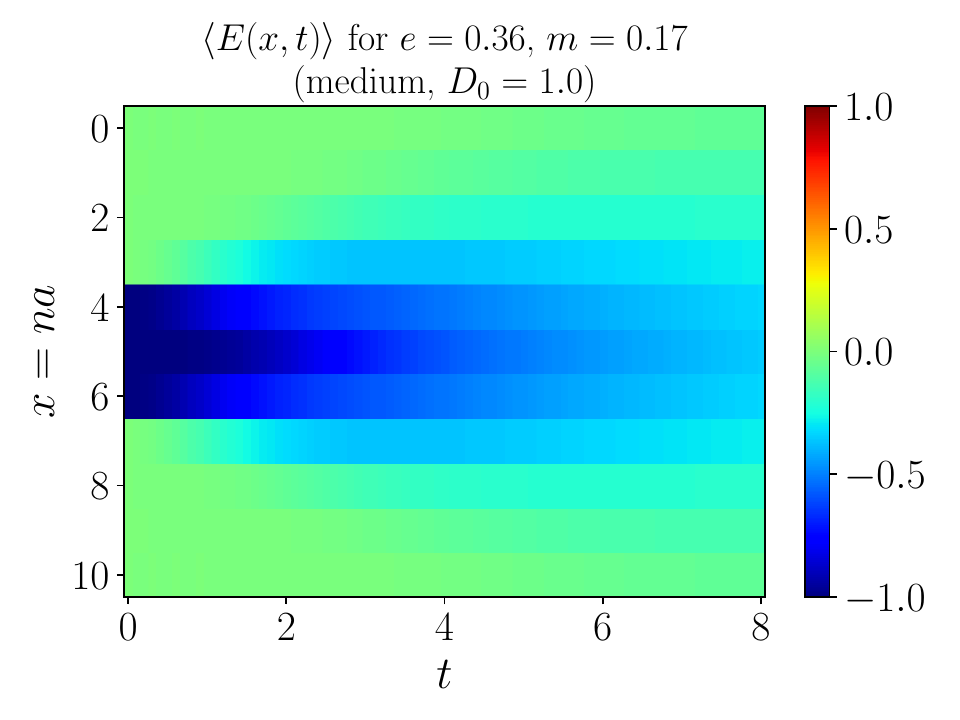}%
}\hfill
\subfloat[\label{fig:string_e}]{%
  \includegraphics[height=1.7in]{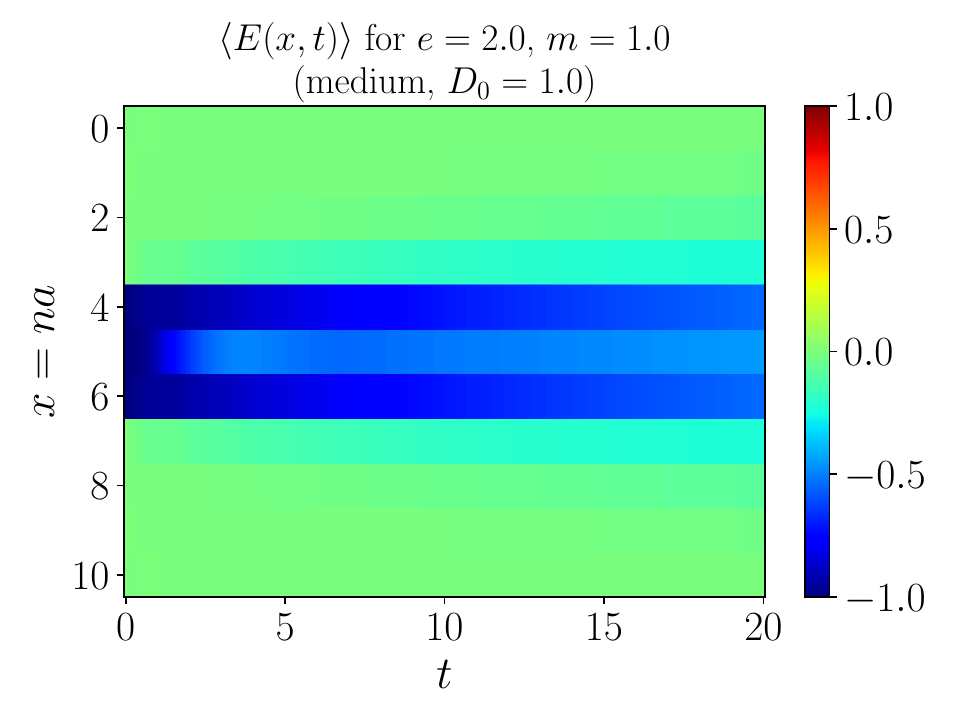}%
}\hfill
\subfloat[\label{fig:string_f}]{%
  \includegraphics[height=1.7in]{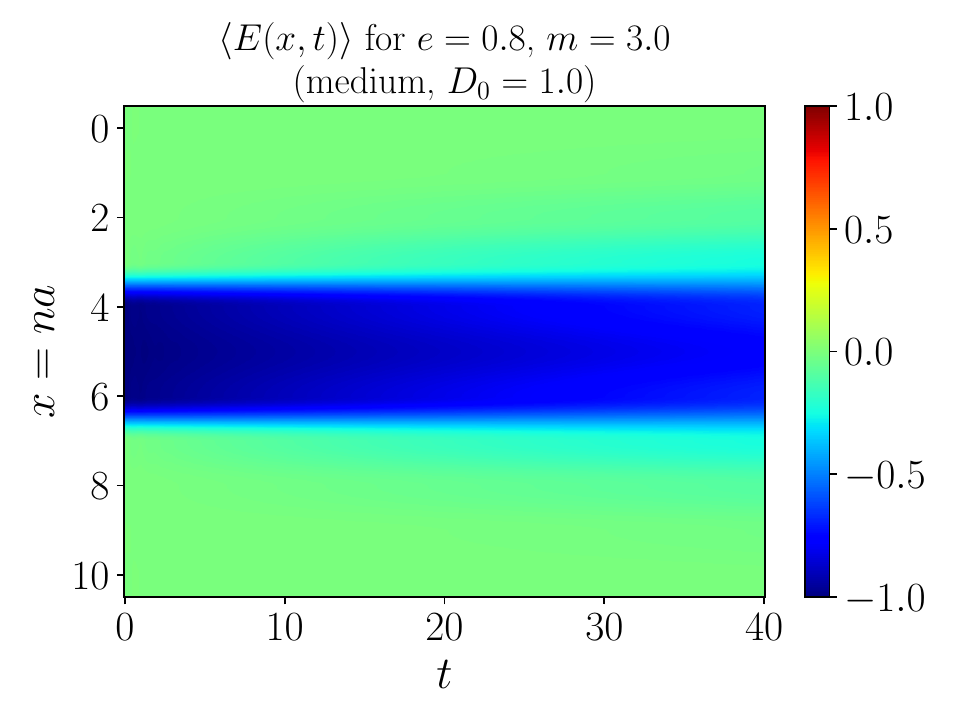}%
}
\caption{Real-time string dynamics in both vacuum (a,b,c) and the medium (d,e,f) with three different sets of parameters of the Schwinger model. For the in-medium evolution, we use the delta correlator with $D_0=1$ and $\beta=0.1$.}
\label{fig:string}
\end{figure*}
\begin{itemize}
    \item In the parameter region where $m\sim e^2\lesssim1$, the fermion mass, the electric energy stored in the electric fields, and the kinetic energy are all on the same order, and string breaking can happen in real-time dynamics. The electric flux between an electron-positron pair will break to release enough energy to create another electron-positron pair to form two charge-neutral mesons that move away from each other. This process can only happen when the electric energy stored in the electric flux is comparable to the sum of the typical kinetic energy and twice the fermion mass. In Fig.~\ref{fig:2D-m-e_vac}, we observe that at small $m$ and $e$ the string breaks, corresponding to large (i.e. less negative) $\overline{E}$. The typical vacuum real-time dynamics in this regime are plotted in Fig.~\ref{fig:string_a}. We predict that states with long initial strings would undergo multiple string breakings, as anticipated from string fragmentation, resulting in the creation of multiple mesons. For instance, if the initial string is $7a$ long, the string can break in three places to produce four mesons. The phenomenon of multiple breakings requires the initial string to be long, which is not considered here.
    \item In the region where $2m\approx e^2 \gg 1$, the string cannot really break in real-time dynamics since the energy released from the breaking of the electric flux is exactly equal or very close to twice the fermion mass, leaving little kinetic energy for the two mesons to carry. The two mesons stay together and after a certain time, the fermion and anti-fermion next to each other convert back into a string. The real-time dynamics is just an oscillation between these two states, i.e. fermion pairs are created and annihilated but the string effectively stays in place and the created mesons do not move away from each other, behaving like a molecule state. The typical vacuum real-time dynamics in this regime are plotted in Fig.~\ref{fig:string_b}. This regime corresponds to the ``wing'' structure in Fig.~\ref{fig:2D-m-e_vac}. It appears as light-blue in the plot, corresponding to an intermediate value of $\overline{E}\approx0.6$ due to configurations where the electric field values on central sites oscillate and the fermion pair is created and annihilated. The period of these oscillations depends on the values of $m,e$ which give rise to a varying magnitude of $\overline{E}$ due to the fixed time window we examine. 
    \item In the region where $m\gg e^2,1$ or $e^2 \gg m,1$, the string stays intact since the energy released from the breaking of one unit electric flux is either too small to create an electron-positron pair in the case with $m\gg e^2,1$, or too large to have kinetic energies of mesons sustainable on the current finite lattice setup in the case with $e^2 \gg m,1$. Processes that significantly violate energy conservation cannot occur at any nonzero time in real-time dynamics. The first case arises due to the inability to create a new electron-positron pair, while the second case is the result of the inability to access states of higher momentum. This regime corresponds to $\overline{E}\approx-1$ (the dark blue region) in Fig.~\ref{fig:2D-m-e_vac} and its typical vacuum real-time dynamics are plotted in Fig.~\ref{fig:string_c}, where the string stays intact during the time evolution. We expect that the string remaining intact in the $e^2 \gg m,1$ region is an artifact of the finite lattice setup that we are studying, which imposes a cutoff on the highest (lowest) momentum state available as $\sim 1/a$ ($\sim 1/(Na)$). In the continuum and infinite volume limits, the system can sustain arbitrary momentum and the energy being released from the string breaking can be converted into kinetic energies. 
\end{itemize}

The modification of these three regimes due to medium effects is shown in Fig.~\ref{fig:2D-m-e_med}, where $\overline{E}$ is obtained from the Lindblad equation as a function of $e,m$. We choose $\beta=0.1$ and the environment correlator to be a delta function with $D_0=0.15$. While we observe the same three regimes as in vacuum, their behaviors are significantly modified. At small $m$ and $e$, we observe a regime of string breaking with slightly larger magnitude of the string flux than in vacuum, due to the delayed breaking effect discussed above. In Fig.~\ref{fig:string_d}, we plot the real-time dynamics in this regime for the medium case with the delta correlator, $D_0=1$ and $\beta=0.1$. We clearly see the quantum dissipation effect caused by the medium, which damps the kinetic energies of the mesons and protects the string from completely breaking. This phenomenon has already been noted in open quantum system studies for quarkonium inside the QGP~\cite{Akamatsu:2018xim,Miura:2019ssi} and may be partially interpreted as quarkonium recombination, a phenomenon known for a long time in the heavy ion community~\cite{Thews:2000rj,Andronic:2003zv,Andronic:2007bi}. At thermal equilibrium, quarkonium dissociation and recombination reach detailed balance~\cite{Yao:2017fuc,Yao:2020xzw}. Similarly, the significant kinetic dissipation observed here can be interpreted as string reconnection in the medium. States with different string configurations in the Schwinger model reach detailed balance when the system thermalizes, driven by the interaction with the thermal environment.

Next, we consider the case where the string does not break in vacuum, i.e. it is a bound state, which happens at larger values of $e,m$. The medium can induce melting of the string, no matter whether the string is oscillating or stable in vacuum, as shown in Figs.~\ref{fig:string_e} and~\ref{fig:string_f} where we use the delta environment correlator again with $D_0=1$ and $\beta=0.1$. The evidence of the medium-induced string breaking can be seen from the lighter blue regions at late times in the center of the lattice. This scenario is analogous to quarkonium dissociation inside a QGP. The medium-induced string breaking rate depends on the parameters of the Schwinger model, as well as the environment correlator. Here we see the string breaking rate is larger for $e=2,m=1$ than $e=0.8,m=3.0$.

\section{Toward quantum simulations: Estimation of Trotter errors~\label{sec:QuantumSimulation}}

\begin{figure}[t]
\includegraphics[scale=0.48]{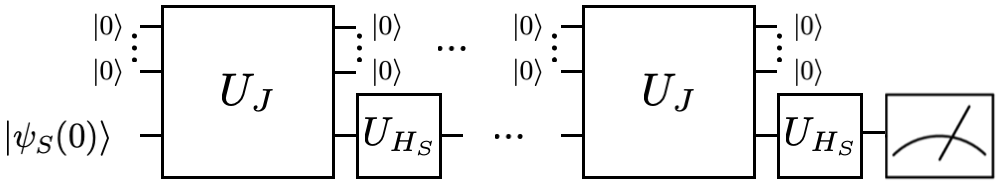}
\caption{Quantum algorithm to simulate Lindblad evolution based on the Stinespring dilation theorem~\cite{nielsen_chuang_2010}. Here $|\psi_S(0)\rangle$ denotes the initial state of the system.~\label{fig:quantum_alg}}
\end{figure}

\begin{figure}[t]
\includegraphics[scale=0.5]{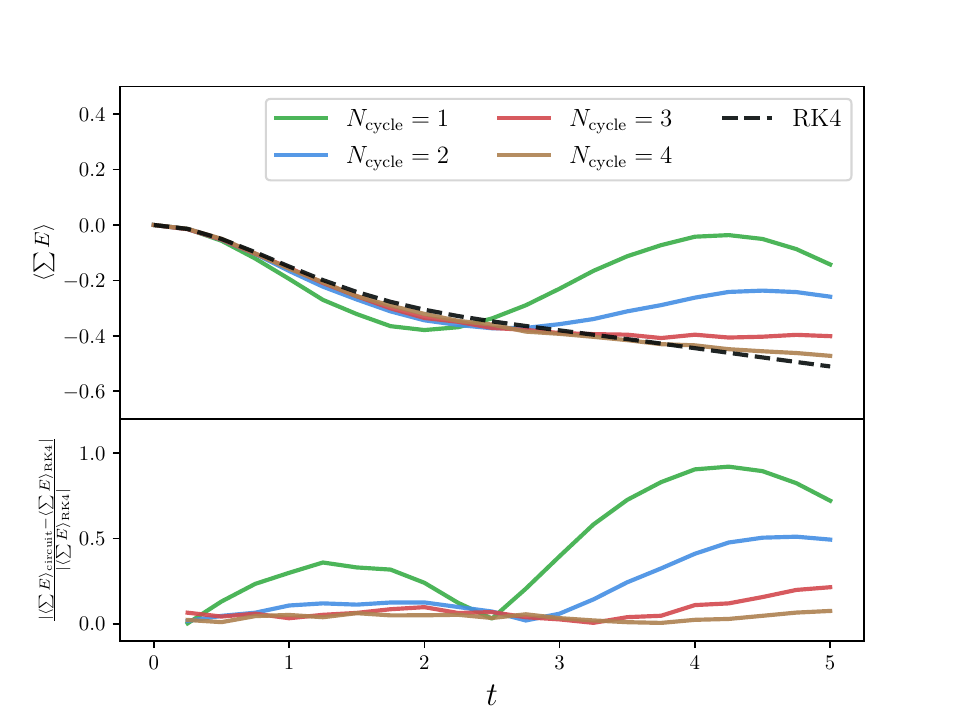}
\caption{Upper panel: Comparison of the full Lindblad evolution (RK4) and results from the (noiseless) quantum simulation using different numbers of cycles $N_{\rm cyl}$, as shown in  Fig.~\ref{fig:quantum_alg}. Lower panel: Ratio of the different approximate results to the full RK4 solution.~\label{fig:Ncycle}}
\end{figure}

Lindblad dynamics can be simulated with a quantum algorithm based on the Stinespring dilation theorem~\cite{nielsen_chuang_2010}. The non-unitary evolution of the open quantum system can be achieved by including an ancillary register, which allows for the embedding of the evolution in an enlarged Hilbert space. In this larger Hilbert space, the evolution is step-wise unitary and repeated reset operations of the ancillary qubit register lead to a time irreversible and non-unitary evolution. Following Ref.~\cite{cleve2016efficient}, we can simulate the Lindblad evolution in terms of small time steps $\delta t=t/N_{\rm cyl}$, where $t$ is the final time we evolve to and $N_{\rm cyl}$ is the number of time steps or cycles. The simulation protocol illustrated in Fig.~\ref{fig:quantum_alg} proceeds by alternating between the application of the unitary evolution operator associated with the system $U_{H_S}=\exp(-iH_S \delta t)$ and the evolution operator $U_J=\exp(-iJ\sqrt{\delta t})$, where $J$ is a block matrix that contains the Lindblad operators in the first row and column
\begin{equation}
    J = \begin{pmatrix}
        0 & L_1^\dagger & \dots & L_m^\dagger\\
        L_1 & 0 & \dots & 0\\
        \vdots & \vdots &\ddots &\vdots\\
        L_m & 0 &\dots & 0
\end{pmatrix} \,.
\end{equation}
Here we limit ourselves to the case where the environmental correlator is given by $D_\delta(x)=\delta_{0x}$, as in the discussion around Eqs.~(\ref{eq:lindblad}) and~(\ref{eq:D}) above. The evolution operator $U_J$ acts on the system and the ancillary register of qubits. The ancillas are reset after every time step $\delta t$, which leads to a non-unitary evolution. In the limit $N_{\rm cyl}\to\infty$, the exact Lindblad evolution is recovered. The error associated with the decomposition of the Lindblad evolution in terms of $U_{H_S}$ and $U_{J}$ operators scales as $\delta t^{1.5}$. To illustrate the numerical size of the error that is introduced by using a finite number of $N_{\rm cyl}$ time steps, we show the evolution of the Schwinger model as an open quantum system for $N_{\rm cyl}=1-4$ in Fig.~\ref{fig:Ncycle} along with the full result based on the fourth order Runge-Kutta (RK4) method. As an example, we consider the expectation value of the electric field in units of $e$ summed over all links of the lattice 
\begin{equation}
\Big\langle \sum_n E(x=na) \Big\rangle \,,   
\end{equation}
with the bare vacuum as the initial state. The same quantum algorithm considered in this section can also be directly used to simulate string breaking or the von Neumann entropy studied in previous sections. All numerical results presented in this section are based on an $N=2$ spatial lattice with $e=0.8$, $m=0.5$, $a=1$, and $\beta=0.1$. 

\begin{figure}[b]
\includegraphics[scale=0.5]{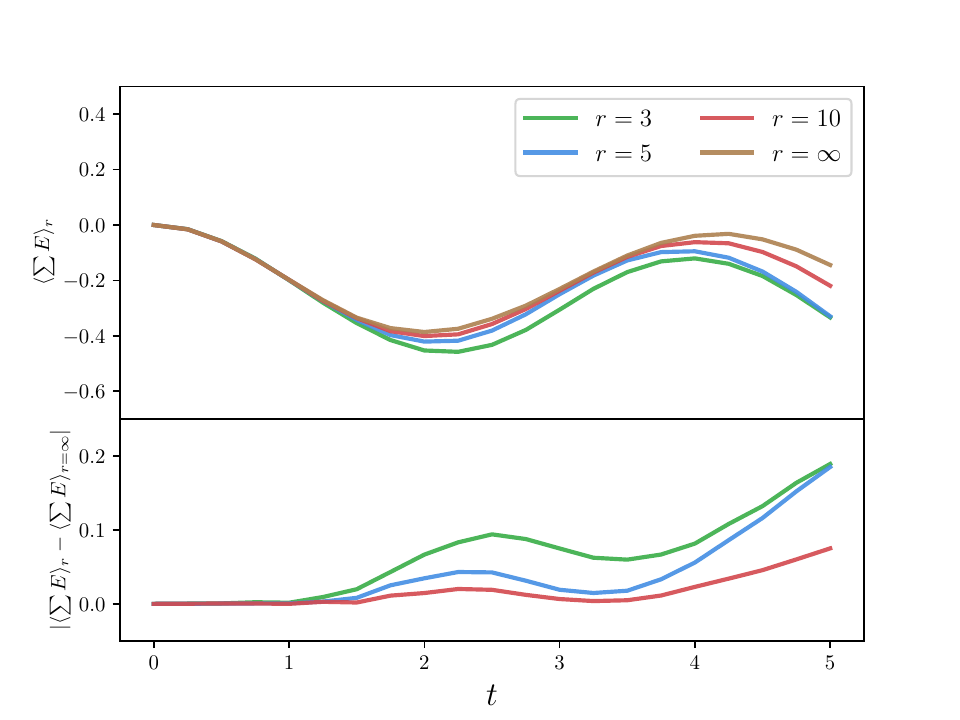}
\caption{Numerical results for the time evolution of the Schwinger as a closed system. Results with a different number of Trotter steps $r$ are shown in comparison to the full result, which is labeled as $r=\infty$.~\label{fig:trotter_vac}}
\end{figure}
As the number of steps $N_{\rm cyl}$ is increased, the agreement with the full result improves. Here we assumed that both $U_J$ and $U_{H_S}$ can be mapped exactly to elementary quantum gates, without considering shot noise and gate errors. In general, the mapping of the unitary evolution operators $U_J$ and $U_{H_S}$ to elementary quantum gates requires further approximations. In Ref.~\cite{DeJong:2020riy,deJong:2021wsd} an efficient compilation method~\cite{2020arXiv201000215H} was used to approximately map the unitary operators $U_J$ and $U_{H_S}$ to elementary quantum gates. However, for unitary operators acting on a larger number of qubits this compilation process can become computationally expensive. Instead, to implement the evolution operators $U_{H_S}$ and $U_{J}$ on a quantum computer, a Trotter-Suzuki decomposition~\cite{BOGHOSIAN199830,doi:10.1126/science.273.5278.1073} can be employed for both operators. This decomposition will introduce additional errors, besides the errors arising due to a finite number of cycles, as shown in Fig.~\ref{fig:quantum_alg}. In this section, we will quantitatively assess both types of errors.

\begin{figure}[!t]
\includegraphics[scale=0.5]{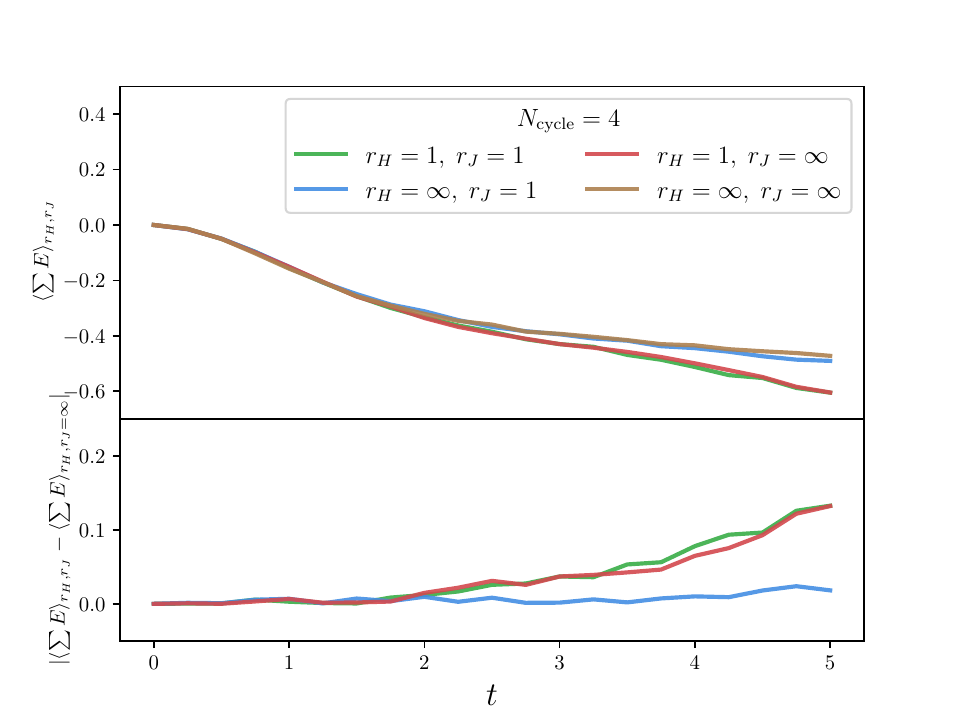}
\caption{Upper panel: Lindblad evolution of the Schwinger model using the quantum algorithm shown in Fig.~\ref{fig:quantum_alg} for $N_{\rm cyl}=4$ as in Fig.~\ref{fig:Ncycle}. We show the result for different numbers of Trotter steps of the operators $U_{H_S}$ and/or $U_J$ indicated by $r_{H,J}$, respectively. Lower panel: Difference between the different Trotter approximations and the result without further Trotter decomposition $r_{H,J}=\infty$.}
\label{fig:trotter}
\end{figure}

We can write any Hamiltonian acting on $n$ qubits, in our case $H_S$ and $J$, as
\begin{equation}\label{eq:HaP}
    H=\sum_j H_j = a_j P_j \,,
\end{equation}
where $P_j\!:\{\mathds{1},X,Y,Z\}^{\otimes^n}$ are strings of $n$ Pauli operators (and the identity). The relevant coefficients $a_j$ can be obtained as
\begin{equation}
    a_j=\frac{1}{2^n}\text{tr}[P_j H] \,.
\end{equation}
The unitary evolution with any of the terms in Eq.~(\ref{eq:HaP}), i.e. $e^{-i H_j t}$, can be directly mapped to elementary quantum gates without further approximations~\cite{doi:10.1126/science.273.5278.1073}. We can implement the evolution of the full Hamiltonian $H$ in Eq.~(\ref{eq:HaP}), using a first order Trotter decomposition
\begin{equation}
    U_1(t)=\prod_{j} e^{-iH_j t}
\end{equation}
The upper bound for the error of this approximation, i.e. the difference between $U_1(t)$ and $e^{-iHt}$, is given by~\cite{PhysRevX.11.011020}
\begin{equation}
    ||e^{-iHt}-U_1(t)|| \leq \frac12 \sum_{j>k} ||[H_j,H_k]||t^2 \,,
\end{equation}
where $||\cdot||$ denotes the spectral norm. The error bound of the first order Trotter decomposition is proportional to the square of the time $t$ and the size of the prefactor depends on the number of non-commuting terms in Eq.~(\ref{eq:HaP}). By decomposing the interval $t$ into $r$ time steps, the error can be reduced to
\begin{equation}\label{eq:trotter_r}
    ||e^{-iHt}-U_1^r(t/r)|| \leq \frac12 \sum_{j>k} ||[H_j,H_k]||\frac{t^2}{r} \,.
\end{equation}
Here $U_1^r(t/r)$ denotes $r$ applications of the Trotter decomposed unitary $U_1$ with each time step being $t/r$. This implies that the value that needs to be chosen for $r$ and the computational cost to perform the simulations within spectral-norm error $\epsilon$, scales as $O(t^2/\epsilon)$. To further reduce the cost, higher order Trotter formulas can be used~\cite{PhysRevX.11.011020,Wecker_2014,Stryker:2021asy}. Here we limit ourselves to first-order Trotter decompositions since we are primarily interested in the difference between the closed and open system evolution.

In the following, we present numerical results for the Lindblad evolution of the Schwinger model using different numbers of Trotter steps. For comparison, we start by considering the error induced by the Trotter decomposition for the vacuum evolution (i.e. $N_{\rm cyl}=1$) of the Schwinger model, which was also considered in Ref.~\cite{Klco:2018kyo}. The results are shown in Fig.~\ref{fig:trotter_vac}, where we choose exemplary values for the number of Trotter steps $r=3,5,10$. As expected, the error increases for late times in comparison to the full result, which is labeled as $r=\infty$. Next, we consider the Trotter error for the Lindblad evolution. Our numerical results are shown in Fig.~\ref{fig:trotter}. For all results, we choose $N_{\rm cyl}=4$, which provides a good approximation of the full result for the time values shown here, as demonstrated in Fig.~\ref{fig:Ncycle}. We denote the number of Trotter steps for $U_{H_S}$ and $U_J$ by $r_{H,J}$, respectively. These Trotter steps correspond to a further decomposition of the time interval of each cycle of time length $\delta t$. In other words, here $r_{H,J}=1$ is analogous to a Trotter decomposition of the vacuum result in Fig.~\ref{fig:trotter_vac} with $r=4$. Interestingly, we observe that the error induced by the Trotter decomposition for the open quantum system evolution is smaller compared to the time evolution of the closed system. This holds even though more qubits and gates have to be applied to realize the Lindblad evolution due to the unitaries $U_J$. For the closed system evolution, we use 3 qubits and for the open system, we need twice as many. From the upper bound for the Trotter error in Eq.~(\ref{eq:trotter_r}) and the sequential application of quantum gates, one might have expected an increased error for the open quantum system evolution as there are significantly more non-commuting terms that contribute to the total error when $U_J$ is included. While these results may not be universally applicable, they suggest the presence of interesting error cancellation effects in the Trotter decomposition of field-theoretical open quantum systems, which motivates further detailed studies in future work. Furthermore, we note that the Trotterization errors associated with $U_J$ are much smaller than those associated with $U_{H_S}$, as shown in Fig.~\ref{fig:trotter}.

\section{Conclusions~\label{sec:Conclusions}}

In this work, we considered the Schwinger model as an open quantum system and studied its Liouvillian dynamics focusing in particular on the string breaking mechanism. This was achieved by coupling the Schwinger model to a thermal environment and in the quantum Brownian motion limit its time evolution is described by a Lindblad evolution equation. We were thus able to extend previous studies of the static string tension in a thermal medium to the dynamical case and observed a delay in the breakup process with a lower relative velocity of the fragments due to kinetic dissipation. We explored the dependence of this effect on system parameters and we also identified regions of medium-induced breaking and reconnection of the string. Due to similarities of the string breaking process in the Schwinger model and QCD hadronization, our results may provide guidance for constructing hadronization models with or without medium~\cite{Han:2016uhh, Fries:2008hs,AbdulKhalek:2021gbh} and help us to decipher real-world collider events. With also significant developments in real-time simulation of scatterings in quantum field theories~\cite{Jordan:2012xnu,Jordan:2011ci,Jordan:2017lea} and simulation of jet production in the Schwinger model~\cite{Florio:2023dke,Barata:2022wim,Delgado:2022snu,Pires:2020urc}, we hope these advancements combined with our work will provide a promising outlook to simulate and study real-time hadronization processes using simulations of real collider scattering. In addition, we studied Liouvillian eigenvalues and eigenmodes for short- and long-range correlated environments. In particular, we studied the late time relaxation dynamics in terms of the von Neumann entropy, which is governed by the Liouvillian gap. We observed that the CP symmetry of the Lindblad equation plays a critical role when the environmental correlator is taken to be a constant. These results set the stage for future investigations such as non-equilibrium phase transitions in quantum field theories. Lastly, we estimated Trotter errors relevant to quantum simulations of open quantum systems. These errors turned out to be relatively small, making simulations of open quantum systems an attractive candidate for the intermediate-term future application of quantum computing.

\begin{acknowledgements}
We thank Adrien Florio, Bert de Jong, Joshua Lin, Di Luo, Giuseppe Magnifico, Ian Moult, Duff Neill, Mateusz Ploskon, Martin Savage, Bjoern Schenke, Stella Schindler, Phiala Shanahan, George Sterman, Raju Venugopalan, and Brayden Ware for helpful discussions. K.L. and X.Y. were supported by the U.S.~DOE under contract number DE-SC0011090. JM is supported by the U.S. Department of Energy, Office of Science, 
Office of Nuclear Physics, under the contract DE-AC02-05CH11231. F.R. is supported by the DOE with contract No.~DE-AC05-06OR23177, under which Jefferson Science Associates, LLC operates Jefferson Lab, and in part by the DOE, Office of Nuclear Physics, Early Career Program with contract No~DE-SC0024358. X.Y. was supported in part by the U.S. Department of Energy, Office of Science, Office of Nuclear Physics, InQubator for Quantum Simulation (IQuS) (https://iqus.uw.edu) under Award Number DOE (NP) Award DE-SC0020970 via the program on Quantum Horizons: QIS Research and Innovation for Nuclear Science.
\end{acknowledgements}

\appendix
\section{Dependence of the Thermalization Rate on the Environment Correlation and System Size~\label{app}}

In the numerical studies presented in the main text, we observe the first Liouvillian gap $\Delta_1$ decreases as the width of the Gaussian environment correlation function increases, and it also decreases with the system size (when the environment correlation is a delta function). Here we provide an analytic explanation.

We perform Fourier transforms in the Lindblad equation~\eqref{eq:lindblad} by introducing
\begin{align}
D(x_1-x_2) &= \frac{1}{N_f}\sum_{k=0}^{N_f-1} D(k) e^{i2\pi k(x_1-x_2)/N_f} \,,\nn\\
L(k) &= \sum_{k=0}^{N_f-1} L(x) e^{-i2\pi kx/N_f} \,,
\end{align}
and then obtain
\begin{align}
\frac{\diff \rho_S(t)}{\diff t} &= -i \big[H_S,  \rho_S(t) \big] + \frac{a^2}{N_f}\sum_{k=0}^{N_f-1} D(k) \nn\\ 
& \times \big(L(k)\rho_S L^\dagger(k) -\frac12 \{L^\dagger(k)L(k),\rho_S\}\big) \,.
\end{align}
The anticommutator part of the Liouvillian operator can be thought of as an imaginary Hamiltonian. We can use it to estimate the relaxation rate of the system:
\begin{align}
\label{eqn:app:gamma}
\Gamma\sim \frac{a^2}{2N_f} \sum_{k=0}^{N_f-1} D(k) L^\dagger(k) L(k) \,.
\end{align}
The operator $L^\dagger(k) L(k)$ is positive semi-definite. As a result, the relaxation rate is increased when the values of $D(k)$ are larger with fixed $N_f$. For example, if we consider a Gaussian environment correlation with width $\sigma$ in position space, which corresponds to another Gaussian with width $1/\sigma$ in momentum space, the contributions to $\Gamma$ from terms with nonzero $k$ are more suppressed as $\sigma$ increases. This is why the Liouvillian gaps characterizing the relaxation rates decrease as the width $\sigma$ becomes larger in Fig.~\ref{fig:deltalimit}.

Next, we consider $D_\delta(x)=\delta_{0x}$ and discuss why the first Liouvillian gap decreases with the system size $N_f$. With this delta correlation function, we have $D_\delta(k)=1$ for all $k$. Whether Eq.~\eqref{eqn:app:gamma} decreases with $N_f$ is not obvious, since each value of $k$ in the summation contributes and there is an overall prefactor $1/N_f$. To understand the $N_f$ dependence of the Liouvillian gaps, we need to analyze the dissipation rate on a deeper level. We consider how an arbitrary eigenstate $|E_n\rangle$ of the system with eigenenergy $E_n$ dissipates by setting $\rho_S=|E_n\rangle \langle E_n|$ and sandwiching the right hand side of Eq.~\eqref{eqn:app:gamma} between $\langle E_n|$ and $|E_n\rangle$. The dissipation rate of this eigenstate is roughly given by
\begin{align}
\label{eqn:app:expA}
\Gamma_n\sim\ & \frac{a^2}{N_f} \sum_{k=1}^{N_f-1} D(k) \nn\\
& \times \big( \langle E_n| L^\dagger(k)L(k) |E_n\rangle - |\langle E_n|L(k)|E_n\rangle |^2 \big) \nn\\
=\ & \frac{a^2}{N_f} \sum_{k=1}^{N_f-1} D(k) \sum_{m\neq n} \langle E_n | L^\dagger(k) | E_m\rangle \langle E_m | L(k) | E_n \rangle \,,
\end{align}
where we have inserted a complete set of eigenstates (for simplicity, we assume no degeneracy in the following). It is worth noting that for each value of $k$, diagonal matrix elements of $L(k)$ do not contribute to Eq.~\eqref{eqn:app:expA}. The typical absolute value of the off-diagonal matrix element $|\langle E_m | L(k) | E_n \rangle|$ is expected to decrease exponentially with the system size for our system (e.g. as in the eigenstate thermalization hypothesis for non-integrable systems). On the other hand, the number of terms in the summation of Eq.~\eqref{eqn:app:expA} is also exponential in the system size, i.e., $\sim e^{cN_f}$ for some constant $c$. However, not all of them are nonvanishing. In fact, only eigenstates $|E_m\rangle$ whose momenta differ from that of $|E_n\rangle$ exactly by $k$ contribute. Their number is still exponential in the system size, by only a fraction of the total number of eigenstates, i.e., $\sim\frac{1}{N_f}e^{cN_f}$, since this fraction is roughly given by the inverse of the number of momentum sectors in the system, which is $N_f$. This explains why the relaxation rate decreases with $N_f$. But it does not explain why the first Liouvillian gap decreases in a specific power law $N_f^{-\alpha}$. It is expected that details of the system may influence the power exponent and we leave a more complete explanation to future work.

\bibliographystyle{utphys}
\bibliography{main.bib}

\end{document}